\documentclass[pdflatex,sn-mathphys-num,iicol]{sn-jnl}

\usepackage{graphicx}%
\usepackage{multirow}%
\usepackage{amsmath,amssymb,amsfonts}%
\usepackage{amsthm}%
\usepackage{mathrsfs}%
\usepackage[title]{appendix}%
\usepackage{xcolor}%
\usepackage{textcomp}%
\usepackage{manyfoot}%
\usepackage{booktabs}%
\usepackage[ruled,vlined,linesnumbered]{algorithm2e}
\usepackage{algorithmicx}%
\usepackage{algpseudocode}%
\usepackage{listings}%
\usepackage{graphicx}   
\usepackage{tikz}       

\usetikzlibrary{arrows.meta, positioning}  

\theoremstyle{thmstyleone}%
\theoremstyle{thmstyletwo}%
\newtheorem{remark}{Remark}%

\theoremstyle{thmstylethree}%

\begin{document}
\title[RL Methodology for COVID-19]
{On a Reinforcement Learning Methodology for Epidemic Control, with application to COVID-19}

\author*[1]{\fnm{Giacomo} \sur{Iannucci}}\email{giacomo.iannucci@ucl.ac.uk}

\author[2]{\fnm{Petros} \sur{Barmpounakis}}\email{pb891@cam.ac.uk}

\author[1]{\fnm{Alexandros} \sur{Beskos}}\email{a.beskos@ucl.ac.uk}


\author[3]{\fnm{Nikolaos} \sur{Demiris}}\email{nikos@aueb.gr}


\affil*[1]{\orgdiv{Department of Statistical Science}, \orgname{University College London}, 
  \orgaddress{\street{Torrington Place}, \city{London}, \postcode{WC1E 7HB}, \country{United Kingdom}}}

\affil[2]{\orgdiv{Department of Oncology}, \orgname{University of Cambridge}, 
  \orgaddress{\street{Hills Road}, \city{Cambridge}, \postcode{CB2 0QQ}, \country{United Kingdom}}}

\affil[3]{\orgdiv{Department of Statistics}, \orgname{Athens University of Economics and Business}, 
  \orgaddress{\street{Patission 76}, \city{Athens}, \postcode{10434}, \country{Greece}}}


\abstract{This paper presents a real-time, data-driven decision‐support framework for epidemic control. We combine a suitably tailored epidemic model with sequential Bayesian inference and reinforcement‐learning (RL) controllers that adaptively choose intervention levels to balance disease burden, such as Intensive Care Unit (ICU) load, against socio-economic costs. We develop a context‐specific cost estimation method via empirical experiments and expert feedback. Two RL policies are implemented: an ICU‐threshold rule carried out via Monte Carlo grid search and a policy involving a posterior‐averaged $Q$‐learning agent. We validate the framework by fitting the epidemic model on publicly available  ICU real data from the COVID-19 pandemic in England and then generating roll-out scenarios under each RL controller allowing to contrast the effectiveness of the RL approaches against the actual government strategy. Over a 300-day COVID-19 period and across varying cost parameters, both controllers substantially reduce ICU burden compared to historical interventions, demonstrating the power of integrating Bayesian sequential learning with RL for optimisation of epidemic control policies.}


\maketitle

\section{Introduction}
\label{sec:intro}

\textbf{Motivation.} The COVID-19 pandemic starkly illustrated that the \emph{timing} and \emph{severity} of intervention during an epidemic is a knife-edge decision: non-pharmaceutical interventions (NPIs) such as stay-at-home orders, school closures and gathering bans can effectively reduce transmission \cite{hsiang2020effect, Barmpounakis2024}, yet NPIs may also inflict steep socio-economic costs, including on education, healthcare and the global economy \cite{nicola2020socioeconomic}. Overly timid measures risk overwhelming hospitals and outstrip Intensive Care Unit (ICU) capacities \cite{schnyder2022rational}, while excessively aggressive lock-downs can stall economic activity and undermine long-term public compliance \cite{gershon2020managing}. Thus, policy-makers must balance the risk of ICU overflow against the escalating social and economic costs of restrictions in real time, all under deep uncertainty regarding disease transmissibility and intervention effectiveness.

\medskip
\textbf{Framework.} To address the above-posed challenges, we cast the epidemic control as a Markov decision process (MDP) problem, built upon a tailor-made compartmental stochastic epidemic model 
which incorporates vaccination roll-outs and their effect in addition to standard compartments for Susceptible, Exposed, Infectious and Removed (SEIR) sup-populations.
We label our model as SEIR--VU, with `V' referring to vaccinations and `U' to units of intensive care.
The requirement for real-time decision-making is greatly facilitated by utilising the Sequential Monte Carlo (SMC) algorithm of SMC\textsuperscript{2} \citep{ChopinJacobPapaspiliopoulos2013,chopin2020introduction}, with the latter maintaining a rolling posterior over model parameters (e.g., transmission rates) and latent states. At each decision point the SMC filter assimilates the latest ICU data and then deploys one of two Reinforcement Learning (RL) \cite{SuttonBarto2018} controllers: (i) an easily comprehensible ICU-threshold policy algorithmically executed via Monte Carlo grid search \cite{silver2010mc}; (ii) a high-performance posterior-averaged tabular-based policy making use of the $Q$-learning algorithm \cite{WatkinsDayan1992}. Such a dual-option design offers policy-makers a choice between transparency and raw performance, while explicitly accounting (in both cases) for parameter uncertainty.

\medskip
\textbf{Contributions.} Our contributions are multi‐fold. First, we enrich the classical SEIR model \cite{SIR} into a SEIR--VU state‐space system which includes explicit ICU compartments (for both unvaccinated and vaccinated individuals) and multi‐dose vaccination strata with waning immunity \cite{Barmpounakis2022}. The state-space model (equivalently, Hidden Markov Model (HMM)) formulation allows the use of real-time SMC learning algorithms. Model observations correspond to easily accessible, publicly available data. In particular, by considering the daily ICU occupancy (assigned a Negative‐Binomial likelihood \cite{held2005statistical} within our stochastic model) 
as the sole data stream, our approach leverages a consistently reported, widely available data-source, ensuring robustness even in low‐surveillance settings \cite{salje2020burden}, thus making our toolkit applicable to a wide variety of settings, including in cases of countries without data-rich systems. A related model is developed in \cite{sonabend2021non}, where the authors use a 
`multi-type' SEIR model accounting for vaccination status and learn the model parameters through multiple data sources in addition to ICU counts, e.g.~corresponding to information obtained from seroprevalence studies and population-level PCR testing. However, such data-streams are not available for many countries and are also unlikely to be reliable at the start of an epidemic. 

In terms of algorithmic appoaches, we embed our model within the state-of-the-art framework of the SMC\textsuperscript{2} algorithm \citep{ChopinJacobPapaspiliopoulos2013, chopin2020introduction}, which re‐weights and rejuvenates sequentially a cloud of particles representing the posterior of model parameters and hidden states \cite{chopin2020introduction} as each new ICU observation arrives, thus guaranteeing that every decision utilises the most up-to-date, data‐driven belief for the state of the epidemic. 

In addition, we develop a real-time planning engine that selects among four ordinal intervention levels, ranging from no measures to full lockdown \citep{Flaxman2020}, using two SMC\textsuperscript{2}-facilitated policy planners: an easily interpretable ICU‐threshold rule (leveraging Monte Carlo roll-out and grid search) and a posterior‐averaged tabular $Q$‐learner.  Both controllers minimise a unified scalar loss,  balancing between a cumulative ICU burden and an action‐dependent intervention cost. This latter cost is calibrated empirically in our implementations via large‐scale synthetic‐data experiments and refined in consultation with epidemiological experts to capture realistic socio‐economic impacts \cite{nicola2020socioeconomic}. 
SMC\textsuperscript{2} is not a new algorithm within the Monte Carlo community, but to the best of our knowledge, this is the first work to integrate SMC\textsuperscript{2}-based Bayesian learning of epidemic parameters and hidden states with RL decision engines, enabling genuine sequential decision‐making under uncertainty, in contrast to prior studies that evaluate fixed NPIs or rely on offline model estimation \cite{Tildesley2006,Ferguson2020,Hellewell2020}.

We validate our framework by replaying England's COVID-19 ICU trajectory. In more detail, we make use of the historical ICU time-series \cite{NHSEnglandCOVIDHospitalActivity} in England due to COVID-19 infections. On each decision day, our pipeline carries out the following tasks: (i) it assimilates the latest observations from the ICUs into its SMC\textsuperscript{2} filter to update the current posterior; (ii) it selects and implements the next intervention, via either $Q$-learning  or via the threshold policy. 
This loop procedure runs for 300 days, selecting and implementing an intervention/action every 10 days. 
Notice that the above experiment requires generation of `counterfactual' ICU counts, as optimal actions under the considered policies will very often differ from the real action taken during the epidemic. For this purpose, we build a data generator by calibrating our formulated SEIR--VU model over the totality of the sequence of ICU data. Such a generator is used to produce daily ICU counts that closely mirror real‐world data.
Overall, we have a mechanism that allows a direct comparison between the real actions taken in England during the pandemic and the counterfactual outcomes under the actions put forward by our two policy controllers. 
Our numerical results show that our real-time RL framework can lead to much improved decision making during an epidemic.



\medskip

\textbf{Related Work.} Epidemic control has traditionally focused on evaluating a small number of fixed intervention strategies, e.g.~\cite{Tildesley2006, Ferguson2020, Hellewell2020} assess lockdowns and school closures via simulation under offline‐estimated models, while deterministic optimal‐control formulations \citep{LedzewiczSchattler2011, ElhiaRachikBenlahmar2013} solve an one‐shot planning problem. Furthermore, \cite{mccabe2021modelling} uses forecasting projections of COVID-19 patients to estimate the demand for and resulting spare capacity of ICU beds, staff and ventilators under different epidemic scenarios in three European countries (France, Germany and Italy) across the winter of 2020.  Separately, Bayesian filtering methods like SMC\textsuperscript{2} \cite{ChopinJacobPapaspiliopoulos2013} have been deployed for sequential parameter inference in stochastic epidemic models and forecasting based on the inferred transmissibility, with application to COVID-19 in Norway \cite{storvik}.

A handful of recent studies bring RL methodology into the area epidemic policy design under uncertainty. \cite{wan2021momrl} calibrates an SIR model to Chinese COVID-19 case data via closed-form inference, and then carries out offline multi-objective policy search, identifying threshold rules and deep $Q$-network policies on the fixed posterior, without using SMC updates or re-optimising policies as new information arrives. \cite{libin2022drl} trains a model‐free PPO agent on a large meta‐population influenza model in Great Britain, assuming known parameters and focusing on weekly school closures. \cite{reymond2022pareto} learns continuous contact‐reduction schedules for Belgium’s first COVID‐19 wave via Pareto‐conditioned networks but, again, under static, offline‐estimated parameters.

\medskip
\textbf{Setting.} We represent a national‐scale epidemic with a suitably tailored SEIR–VU compartmental model featuring five vaccine‐induced immunity classes and waning protection. Every 10 days, our decision engine, implemented for two choices of policies, selects one of four ordered NPI intensities: \emph{minimal}, \emph{moderate}, \emph{strong} or \emph{very strong}, each corresponding to a specific transmission rate parameter. The controller balances two conflicting objectives: minimising both the cumulative ICU burden and the socio‐economic cost of interventions.  

\medskip
\textbf{Challenge.}  Both the transmission parameters and the
effectiveness of NPIs evolve as pathogen properties, societal behaviour and
vaccine coverage change.  Optimal control therefore requires: (i)
sequential Bayesian inference of the transmission rate parameters; (ii) a
planning algorithm that hedges against that uncertainty.

\medskip
\textbf{Outline.} The remainder of this paper is structured as follows. Section \ref{sec:Pre} covers preliminaries on MDPs, RL and compartmental epidemic models. In Section \ref{sec:1} we present our real-time decision-making framework, illustrating the interactions amongst  the SEIR--VU model, the SMC\textsuperscript{2} filter and the decision engine. Section \ref{sec:MDP} presents the MDP formulation for the problem at hand. Section \ref{decision tool} describes the generation of  intervention strategies by the two planners (an interpretable ICU-threshold policy and a posterior-averaged $Q$-learner). Section \ref{results} reports our numerical experiments on historical ICU data in England, comparing both controllers against the real interventions. We conclude in Section \ref{conclusion} with a discussion of our findings and directions for future work.

\section{Preliminaries} \label{sec:Pre}

\subsection{Markov Decision Process} 
\label{sec:mdp}

A Markov Decision Process (MDP) \citep{puterman2014markov} is specified by the tuple:
\begin{equation*}
\bigl(\mathcal{S},\,\mathcal{A},\,\mathcal{P},\,r,\,\gamma\bigr),
  \label{eq:mdp_tuple}
\end{equation*}
where $\mathcal{S}$ is a finite set of states, $\mathcal{A}$ a finite set of actions, $\mathcal{P}\colon\mathcal{S}\times\mathcal{A}\to
\mathsf{P}(\mathcal{S})$, with $\mathsf{P}(\mathcal{S})$ the set of probability measures on $\mathcal{S}$, is the transition kernel describing the dynamics of a process $\{S_t\}$ so that:
\begin{equation*}
  \mathcal{P}(s'\,|\,s,a) 
    = \Pr\,[\,S_{t+1}=s'\,|\,S_{t}=s,A_{t}=a\,],
  \label{eq:transition_kernel}
\end{equation*}
and $r\colon \mathcal{S}\times\mathcal{A}\to\mathbb{R}$ is the reward function:
\begin{equation}
  r(s,a) 
    = \mathrm{E}\,[\,R_{t}\,|\,S_{t}=s,A_{t}=a\,]
  \label{eq:reward_function}
\end{equation}
that involves the random variable $R_{t}$.
We denote the discount factor by   
$\gamma \in [0,1)$. 
A \emph{policy} $\pi:\mathcal{S}\to \mathcal{A}$ is a mapping from states to actions.  
%
  %
    %

%



\medskip

\textbf{Value Functions.} 
There are two quantities that are central to finding optimal policies: the state value and the state-action value functions \citep{SuttonBarto2018}. These serve as cornerstones for evaluating policies and are key components of RL algorithms, such as $Q$-learning. 
The state value function $V^\pi(s)$ represents the expected discounted cumulative reward when starting from state $s$ and following  policy $\pi$.   Formally, assuming an infinite horizon: 
\begin{equation}
  V^\pi(s)
  := \mathrm{E}^\pi\bigl[\,\sum_{t=0}^{\infty}\gamma^{t}\,r(S_{t},A_{t})
    \bigm|S_{0}=s\,\bigr].
  \label{eq:V_def_explicit}
\end{equation}
The expectation $\mathrm{E}^\pi$ is taken w.r.t.~the trajectory:
\begin{equation*}
  S_{0},\;
  A_{0}= \pi(S_0),\;
  S_{1}\sim \mathcal{P}(\cdot\,|\,S_0,A_{0}),\; \ldots.
\end{equation*}
The action-state value function, $Q^\pi(s,a)$, is the expected cumulative reward when starting in state $S_0=s$, taking action $A_0=a$ and then following policy $\pi$, that is:
\begin{equation*}
  Q^\pi(s,a)
  := \mathrm{E}^\pi\bigl[\,\sum_{t=0}^{\infty}\gamma^{t}\,r(S_{t},A_{t})
    \,\big|\,S_{0}=s,A_{0}=a\,\bigr].
  \label{eq:Q_def_explicit}
\end{equation*}
An overarching goal is to identify an optimal policy 
$\pi^*$ that maximises the expected return from every state, that is, to find a policy under which the state value function (\ref{eq:V_def_explicit}) is as large as possible for all $s \in \mathcal{S}$ \citep{vanotterlo2012rlmdp}. 
\medskip

\textbf{Optimal Policy.}
Note that the above description suggests that an action requires a policy-based decision at each time unit. More generally, the set of decision times can be a subset of the time instances under consideration, as it will be the case with the main application of interest in this work. 
We consider two structures of policies for the actions. 

The one policy puts in place thresholds on the observed scalar sub-component, $Y_t$, of the full state $S_t=(X_t,Y_t)$. This setting corresponds to the  easily interpretable ICU-threshold policy mentioned in Section \ref{sec:intro}. In particular, $Y_t$ refers to the number of occupied ICUs due to COVID-19 infections, whereas $X_t$ denotes the unobserved numbers in the rest of the compartments of our SEIR--VU model. More details will be provided in the sequel. Finding the optimal policy at a decision time, say $t_0$, in this setting involves a `brute-force' optimisation of the state value function (\ref{eq:V_def_explicit}) -- adjusted in the obvious manner so that the initial time instance now being $t_0$ instead of $0$ -- over the thresholding parameters.  

The second direction allows more general structures for the policy $\pi:\mathcal{S}\to \mathcal{A}$. One then obtains the optimal policy via use of the $Q$-learning algorithm \citep{SuttonBarto2018}. Specifically, in this case the mapping $\pi:\mathcal{S}\to \mathcal{A}$ is constructed by splitting the values of $Y_t$ in a number of $G\ge 1$ groups, 
each assigned a label $g\in\mathcal{G}:=\{1,2\ldots, G\}$  (recall that $Y_t$ is the observed number of occupied ICUs).
We thus consider variates $G_t$ representing the group at time $t$, with realisations denoted by $g_t$.
At a decision time $t_0$, $Q$-learning involves the selection of some simple initial tabular values $Q_0(g,a)$, $g\in\mathcal{G}$, $a\in\mathcal{A}$, and then the implementation of the follow iterations, for $k\ge 0$, for  the particular $g\equiv g_{t_0+k}$, $y\equiv y_{t_0+k}$, $g'\equiv g_{t_0+k+1}$ and action $a\equiv a_{t_0+k}$: 
\begin{align}
&Q_{k+1}(g,a) = Q_{k}(g,a)
\label{eq:q_update} + \alpha_{k}\\
&\quad \times \Big[r(y,a) + \gamma \max_{a'}Q_{k}(g',a')
 - Q_{k}(g,a)\Big]. \nonumber 
\end{align}
with $r(y,a)$ being the reward. 
The action $a=a_{t_0+k}$ is obtained by either the optimisation of the current $Q_k(s,\cdot)$, i.e.~$\arg \max_a Q_k(s,a)$ or (with a small probability) in a random fashion along the lines of a `greedy' $Q$-learning approach balancing exploitation and exploration \citep{SuttonBarto2018}.
Also, $\alpha_{k}\in(0,1)$ is a user-specified positive sequence, satisfying the standard 
Robbins-Monro-type conditions $\sum_k\alpha_{k}=\infty$, $\sum_k\alpha_{k}^2<\infty$. 
In the context of our epidemic application, iteration (\ref{eq:q_update}) makes use of simulated states from time $t_0+1$ to $t_0+H$, for a time horizon $H>0$ -- in fact, convergence requires consideration of a multitude, say $E>0$, of such `episodes' over $t_0+1,\ldots, t_0+H$, so the total number of iterations is $H\times E$.  
Once $Q_{k}$ has converged to some $Q^{*}$, the agent can obtain an optimal policy, at $t_0$, via: 
\begin{equation}
\label{eq:optimal_pi}
\pi^*(g) = \arg\max_{a\in\mathcal{A}} Q^*(g,a).
\end{equation}

\begin{remark}
We note that convergence of $Q$-learning iterates is predominantly proven under a Markovian assumption for the state used as an input in the tabular $Q$-function. In (\ref{eq:q_update}), the use of information only about $Y_t$ in the specification of $Q$ implies that such input will typically be non-Markovian in our setting. It is beyond the scope of this work to provide an analytical proof of convergence -- instead we will rely on experimental results to illustrate that iterate (\ref{eq:q_update}) converges to a limit. 
\end{remark}
\subsection{Compartmental Models}
\label{sec:seir}

Compartmental models constitute the most widely used class of stochastic processes for the study of infectious-disease dynamics \citep{Tang2020}.  Herein, the population is partitioned into mutually exclusive `compartments' and the flow of individuals between compartments over time describes the progression of the epidemic.

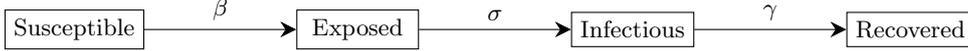
\begin{figure*}[t]
  \centering
  \begin{tikzpicture}[
      font=\small,
      node distance=2cm,
      >={Stealth[length=6pt,width=6pt]}   
    ]
    \node[draw, rectangle, minimum width=1.6cm] (S) {Susceptible};
    \node[draw, rectangle, minimum width=1.6cm, right=of S] (E) {Exposed};
    \node[draw, rectangle, minimum width=1.6cm, right=of E] (I) {Infectious};
    \node[draw, rectangle, minimum width=1.6cm, right=of I] (R) {Recovered};

    \draw[->] (S) -- (E) node[midway, above] {$\beta$};
    \draw[->] (E) -- (I) node[midway, above] {$\sigma$};
    \draw[->] (I) -- (R) node[midway, above] {$\gamma$};
  \end{tikzpicture}
  \caption{Flow diagram of the SEIR model.}
  \label{fig:seir}
\end{figure*}

The classical \emph{Susceptible-Exposed-Infectious-Recovered}
(SEIR) model extends the simpler SIR framework by adding an \emph{Exposed} compartment to capture the incubation period during which individuals are infected but not yet infectious.  Let:
\begin{equation*}
  S(t),\; E(t),\; I(t),\; R(t),
\end{equation*}
denote the numbers of susceptible, exposed, infectious, and recovered (or removed) individuals, respectively, at time $t$, and let $N = S(t) + E(t) + I(t) + R(t)$ be the total (constant) population.  The standard deterministic SEIR model is governed by the ordinary differential equation (ODE):
\begin{align}
  \tfrac{dS}{dt} &= -\beta \,\tfrac{S\,I}{N}, \nonumber \\[0.2cm]
  \tfrac{dE}{dt} &=  \beta \,\tfrac{S\,I}{N} \;-\; \sigma\,E, \nonumber\\[0.2cm]
  \tfrac{dI}{dt} &=  \sigma\,E \;-\; \gamma\,I, \nonumber\\[0.2cm]
  \tfrac{dR}{dt} &=  \gamma\,I,\nonumber
\end{align}
%
where $\beta$ is the transmission rate, $\sigma$ is the progression rate  (incubation) from exposed to infectious;
$\gamma$ is the removal (recovery or death) rate of an infectious individual.
We will ultimately adopt a stochastic formulation of the SEIR model as such an approach better reflects the uncertainty and variability inherent in epidemic dynamics, especially in finite populations \citep{allen2010stochastic}.
A diagram for the SEIR model is
given in Figure \ref{fig:seir}.

In Section~\ref{sec:1} we extend this basic SEIR system by adding vaccination strata, waning immunity and explicit ICU compartments, and embed our model within a state‐space formulation amenable to Bayesian filtering methodologies.

\subsection{Hidden Markov Models}
\label{sec:HMM}

A hidden Markov model (HMM) provides a flexible way to probabilistically relate a latent Markovian process with noisy observations \citep{durbin2012state}.  At each time $t$, we assume:
\begin{align}
  X_{t}\,|\,X_{t-1} &\sim p(X_{t}\,|\,X_{t-1},\theta), 
  \label{eq:ssm_process}\\
  Y_{t}\,|\,X_{t} &\sim p(Y_{t}\,|\,X_{t},\theta),
  \label{eq:ssm_observation}
\end{align}
where $X_t$ is the (unobserved) latent state at time $t$, 
$Y_t$ is the observation and $\theta$ is a vector of static parameters that might govern both the signal and the observation noise.
In our setting, the latent state $X_t$ will represent the SEIR, vaccination and ICU compartments, while $Y_t$ will correspond to daily ICU counts. 
In the MDP terminology of Section \ref{sec:mdp}, the Markovian state $S_t$ consists of both $X_t$ and $Y_t$, i.e.~$S_t = (X_t, Y_t)$.

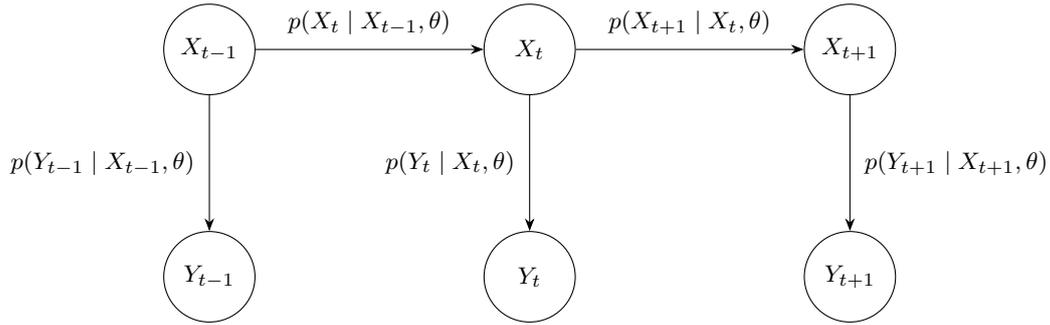
\begin{figure*}[t]
  \centering
  \begin{tikzpicture}[font=\small, node distance=3cm, >=Stealth]
    \tikzstyle{ssmnode} = [draw, circle, minimum size=1.2cm, inner sep=0pt]

    \node[ssmnode] (Xt-1)   {\(X_{t-1}\)};
    \node[ssmnode, right=of Xt-1] (Xt)     {\(X_t\)};
    \node[ssmnode, right=of Xt]   (Xt+1)   {\(X_{t+1}\)};

    \node[ssmnode, below=1.8cm of Xt-1] (Yt-1) {\(Y_{t-1}\)};
    \node[ssmnode, below=1.8cm of Xt]   (Yt)   {\(Y_t\)};
    \node[ssmnode, below=1.8cm of Xt+1] (Yt+1) {\(Y_{t+1}\)};

    \draw[->] (Xt-1) -- (Xt)   node[midway, above, yshift=2pt] {\(\,p(X_t\mid X_{t-1},\theta)\,\)};
    \draw[->] (Xt)   -- (Xt+1) node[midway, above, yshift=2pt] {\(\,p(X_{t+1}\mid X_t,\theta)\,\)};

    \draw[->] (Xt-1) -- (Yt-1) node[midway, left,  xshift=-2pt] {\(p(Y_{t-1}\mid X_{t-1},\theta)\)};
    \draw[->] (Xt)   -- (Yt)   node[midway, left,  xshift=-2pt] {\(p(Y_{t}\mid X_t,\theta)\)};
    \draw[->] (Xt+1) -- (Yt+1) node[midway, right, xshift= 2pt] {\(p(Y_{t+1}\mid X_{t+1},\theta)\)};
  \end{tikzpicture}
  \caption{State‐space model over three time steps. Latent states $X_{t-1},X_t,X_{t+1}$ evolve according to the Markov dynamics \eqref{eq:ssm_process}, and each gives rise to a noisy observation amongst $Y_{t-1},Y_t,Y_{t+1}$ via \eqref{eq:ssm_observation}.}
  \label{fig:ssm_diagram_extended}
\end{figure*}

SMC algorithms can be used to obtain samples from the sequence of posterior distributions
%
$p(\theta, X_{1:t}\,|\,Y_{1:t})$
%
via a population of weighted particles that are re‐weighted, resampled and rejuvenated with each arriving $Y_t$. Such an approach yields a \emph{sequential} Bayesian update of the latent state and parameters, ensuring every decision uses the most current, data‐driven belief.
To update our information about $\theta$ in an sequential 
manner, we will embed our HMM within the framework of the  \emph{SMC\textsuperscript{2}} algorithm \citep{ChopinJacobPapaspiliopoulos2013}.  
Figure~\ref{fig:ssm_diagram_extended} illustrates the evolution of a HMM over three time steps, highlighting the sequential dependence between latent states and observations.


\subsection{SMC\texorpdfstring{\textsuperscript{2}}{2}-Algorithm}
\label{sec:smc2}

We embed our SEIR--VU model within the SMC\textsuperscript{2} framework which targets at every time $t$ the joint posterior of an HMM with latent Markovian state $X_t$ and observation $Y_t$, expressed as follows:
\begin{align*}
p(\theta, x_{0:t}\,|\,&y_{0:t})\propto 
p(\theta)\,p(x_0\mid\theta)\times 
\\ & \prod_{s=1}^{t} p(x_s\,|\,x_{s-1},\theta)\,p(y_s\,|\, x_s,\theta).
\end{align*}
SMC\textsuperscript{2} maintains an \emph{outer} population of weighted parameter particles $\{\theta^{(i)},W_t^{(i)}\}_{i=1}^{N_\theta}$ and, for each such $\theta^{(i)}$, an \emph{inner} particle population over latent states $\{x_{0:t}^{(i,j)},w_t^{(i,j)}\}_{j=1}^{N_x}$, for some $N_\theta, N_{x}\ge 1$. The inner filters provide unbiased  estimates $\hat{p}(y_t\,|\,y_{0:t-1},\theta^{(i)})$ of the incremental likelihood terms $p(y_t\,|\,y_{0:t-1},\theta^{(i)})$, which are used to update the parameter weights, so that:
\begin{align*}
W_t^{(i)} \;\propto\; W_{t-1}^{(i)}\cdot \hat{p}(y_t\,|\, y_{0:t-1},\theta^{(i)}),
\end{align*}
followed by resampling and rejuvenation of $\theta$ via \emph{resample--move} steps (utilising particle-MCMC kernels). 
The output relevant to our sequential decision making setting is an evolving collection of particles representing the posterior $p(\theta, x_{t}\,|\,y_{0:t})$.
In such a manner, every decision point makes use of  a posterior law that has assimilated all data available up to the present day.

At a decision time instance $t_0$, SMC\textsuperscript{2} incorporates the latest ICU data to deliver:
(i) posterior draws of parameters $\{\theta^{(k)}\}_{k=1}^{K}\sim p(\theta\,|\,y_{0:t_0})$; and
(ii) for each $\theta$-draw, samples of latent states from  $p(x_{t_0}\,|\,\theta^{(k)},y_{0:t_0})$ via the inner filters. 
A set of available joint samples  $\{(\theta^{(k)},x_{t_0}^{(k)})\}_{k=1}^{K}$ are called upon to initialise the training roll-outs used by the RL controllers, as described in detail in Sections~\ref{Treshold}--\ref{subsec:tabular} in the sequel.

Note that an SMC\textsuperscript{2} update at time $t$ targets the full posterior $p(\theta,x_{0:t}\,|\,y_{0:t})$, i.e.~likelihood contributions are defined 
w.r.t.~all data from $0$ to $t$, thus SMC\textsuperscript{2} provides a computationally effective sequential approach even if not an `online' one \citep{ChopinJacobPapaspiliopoulos2013}.

\section{SEIR–VU Model}  
\label{sec:1}

We extend the SEIR model discussed in Section \ref{sec:seir} by augmenting its state to include compartments for vaccinated sub-populations and ICU numbers. Simple deterministic SEIR dynamics may not efficiently capture the day‐to‐day variability observed in real epidemic data, so we reformulate the transitions via use of Binomial distributions, a choice popular in the literature for its clear probabilistic interpretation and ability to model stochasticity \citep{held2019}. 

\subsection{Model Compartments}
\label{ssec:compartments}

\begin{figure*}[!t]
  \centering
  \includegraphics[width=0.9\textwidth]{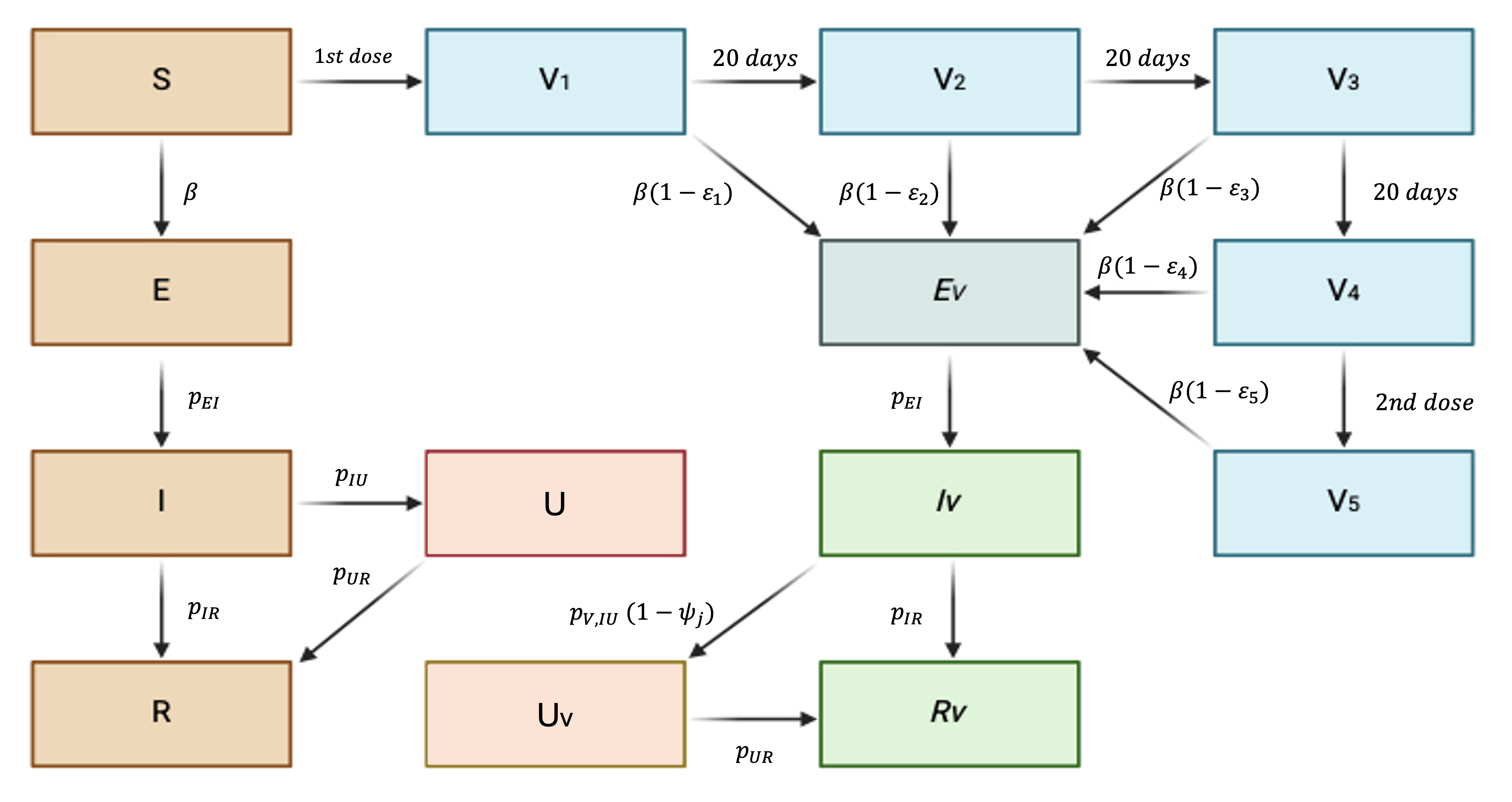}
  \caption{SEIR--VU model dynamics.}
  \label{fig:generalised_seir_model}
\end{figure*}

We develop the 14-compartment stochastic SEIR--VU model to describe the COVID-19 epidemic in the UK, with a population of size $N=68\times10^{6}$. 
On each day $t$, the system state is comprised of the following co-ordinates: 
\begin{align}
X_{t} &= \bigl(S(t),\;E(t),\;I(t),
\;R(t),\;ICU(t),\nonumber \\[0.1cm] &\qquad\,\, V_{1}(t),\;V_{2}(t),\;V_{3}(t),\;V_{4}(t),\;V_{5}(t),\nonumber\\[0.05cm] &\quad \qquad\,\,\, E_{V}(t),\;I_{V}(t),\;R_{V}(t),\;ICU_{V}(t)\bigr).
\label{eq:state}
\end{align}
In (\ref{eq:state}), $S(t)$ denotes susceptible individuals. Then,  $E(t)$, $I(t)$ and $R(t)$ denote (respectively) \emph{unvaccinated} exposed,  infectious and recovered individuals. $ICU(t)$ is the number of unvaccinated individuals in an ICU.
 $E_{V}(t)$, $I_{V}(t)$, $R_{V}(t)$, $ICU_{V}(t)$ are the corresponding compartments for \emph{vaccinated} individuals.
Compartments $V_{j}(t)$ represent vaccinated individuals of immunity level $j$, where $1\le j \le 5$. The sum of counts in all compartments remains fixed at $N$.
Fig.~\ref{fig:generalised_seir_model} gives a schematic overview of all transitions across susceptible, exposed, infectious, recovered, ICU and vaccinated states.

%
\medskip
\textbf{Vaccination Procedure.} 
Daily vaccination numbers in England starting from the late December 2020 are available from NHS reports \citep{NHSEnglandVaccinations}. We denote below by `$\mathrm{daily\_first}_{t}$' and `$\mathrm{daily\_second}_{t}$' the number of first and second COVID-19 vaccinations, respectively, administered to individuals on day $t$ as recorded in official NHS England data.
First-dose vaccinations move individuals from $S(t)$ into the first vaccinated compartment $V_{1}(t)$. Transitions then occur via the path $V_{1}(t)\to V_{2}(t)\to V_{3}(t)\to V_{4}(t)$, unless a second vaccination takes place in which case individuals move to $V_{5}(t)$. Second-dose vaccinations prioritise individuals in $V_{4}(t)$ and then in $V_{3}(t)$. Vaccinated individuals also move to the compartment of vaccinated exposed individuals $E_{V}(t)$. Full details of the involved transition dynamics are provided later in the text.

Table \ref{tab:efficacy} shows the vaccine efficacy for the five vaccination compartments, in terms of the values of immunity levels $\{\varepsilon_j\}$ and protection from ICU admission $\{\psi_j\}$. 
Notice that transition to $V_{5}(t)$ is accompanied by an instantaneous increase in immunity to $\varepsilon_5 = 0.95$ in our model. 
This assumption is supported by immunological studies showing that for the SARS-CoV2 variant second-dose protection builds rapidly due to memory B cells primed by the first dose \citep{Kim2024KineticsCOVIDVaccination}. In contrast, our model assumes that protection from the first dose increases more slowly: individuals are first moved to $V_{1}(t)$ and then gradually acquire higher immunity as they transition to $V_{2}(t)$. This reflects the empirical observation that the first dose requires a delay (typically 10–20 days) before meaningful immune protection is achieved \citep{Chau2022AZImmunogenicity}.

\begin{table}[h]
\centering
\begin{tabular}{c|ccccc}
Level $j$ & 1 & 2 & 3 & 4 & 5 \\
\hline
$\varepsilon_j$ & 0.50 & 0.80 & 0.70 & 0.60 & 0.95 \\
$\psi_j$ & 0.50 & 0.75 & 0.70 & 0.60 & 0.89
\end{tabular}
\caption{Vaccine efficacy by immunity level $\varepsilon_j$ and protection from ICU admission $\psi_j$, for $1\le j \le 5$.}
\label{tab:efficacy}
\end{table}

\subsection{State Transition}
\label{subsec:state}

We describe in full detail the dynamics governing the transitions between the 14 compartments making up state $X_t$ in (\ref{eq:state}) and  shown in Figure~\ref{fig:generalised_seir_model}.

Starting from the susceptible group $S(t)$, we have the transition counts: 
\begin{align*}
SE(t) &\sim \mathrm{Binomial}(S(t),\;p_{SE,t}),\\
SV(t)  &= \mathrm{daily\_first}_t,
\end{align*}
for the probability:
\begin{align}
\label{eq:beta1}
p_{SE,t} = 1 - \exp\big\{-\beta\, \tfrac{I(t)+I_{V}(t)}{N}\big\}.
\end{align}
$SE(t)$ denotes the transitions from $S(t)$ to $E(t)$ in one time unit (a day) and $SV(t)$ the movements from $S(t)$ to $V_{1}(t)$. 
Thus, we have the state transition: 
\begin{align*}
S(t+1) = S(t) - SE(t)  -SV(t).
\end{align*}
We follow similar notation conventions to the ones above in the expressions below.  
%
%
%

\medskip
\textbf{Unvaccinated Compartments.} 
%
For the counts $E(t)\to I(t)$ and the transitions for 
$I(t)\to ICU(t)$, $I(t)\to  R(t)$, $ICU(t)\to  R(t)$ we have: 
\begin{align*}
EI(t) &\sim \mathrm{Binomial}(E(t),\,p_{EI}),\\
IR(t) &\sim \mathrm{Binomial}(I(t),\,p_{IR}),\\
IU(t) &\sim \mathrm{Binomial}(I(t),\,p_{IU}),\\
U\!R(t) &\sim \mathrm{Binomial}(ICU(t),\,p_{U\!R}),
\end{align*}
for probability parameters $p_{EI}$, $p_{IR}$, $p_{IU}$ and $p_{U\!R}$. Thus, the state dynamics write as:  
\begin{align*}
  E(t+1)  &= E(t) - EI(t),   \\                  
  I(t+1)   &= I(t) + EI(t) - IR(t) - IU(t),\\
  U(t+1)   &= U(t) + IU(t) - U\!R(t),\\
  R(t+1) &= R(t) + IR(t) +U\!R(t).
\end{align*}

\medskip
\textbf{Vaccinated Compartments.} 
Taking under consideration the vaccination procedure described earlier, we quantify the prioritisation of groups $V_{3}(t)$ and $V_{4}(t)$ for the purposes of second vaccination dose by setting:  
\begin{align*}
\mathrm{daily\_second}_{4,t} &= \min\big\{\mathrm{daily\_second}_{t},V_{4}(t)\big\},\\
\mathrm{daily\_second}_{3,t} &= \min\big\{ V_{3}(t),\mathrm{daily\_second}_{t} \\ &\qquad \qquad \quad - \mathrm{daily\_second}_{4,t}  \big\}.
\end{align*}
To capture the state transitions $V_{1}(t)\to V_{2}(t)\to V_{3}(t)\to V_{4}(t)$, we define the following counts,  for $1\le j\le 3$: 
\begin{align*}
  VV_{j}(t) &\sim \mathrm{Binomial}(V_{j}(t),\,p_{VV}), 
\end{align*}
for a parameter $p_{VV}$ that reflects the change with time in the protection offered by vaccination.
We also have movements involving the four compartments $E_{V}(t)$, $I_{V}(t)$, $ICU_{V}(t)$, $R_{V}(t)$. We define the counts, for $1\le j\le 5$: 
\begin{align*}
V\!E_{j}(t) \sim \mathrm{Binomial}(V_{j}(t),\,p_{V\!E,j,t})
\end{align*}
where we use the probabilities: 
\begin{align}
\label{eq:beta2}
p_{V\!E,j,t} = \beta\,\tfrac{I(t)+I_{V}(t)}{N}\,(1- \varepsilon_j),
\end{align}
thus quantifying the idea that each vaccinated class $V_{j}(t)$ can become exposed at a reduced rate $(1-\varepsilon_j)$ against the baseline. We also define:
\begin{align*}
  EI_{V}(t) &\sim \mathrm{Binomial}(E_{V}(t),\,p_{EI}),\\
   IR_{V}(t) &\sim \mathrm{Binomial}(I_{V}(t),\,p_{IR}),\\
IU_{V}(t) &\sim \mathrm{Binomial}(I_{V}(t),\,p_{V,IU}),\\
U\!R_{V}(t) &\sim \mathrm{Binomial}(ICU_{V}(t),\,p_{U\!R}),
\end{align*}
for a new  parameter $p_{V,IU}$. The remaining three rates $p_{EI}$, $p_{IR}$, $p_{U\!R}$ are assumed to be the same as for the unvaccinated groups. 
In summary, we have the state transitions: 
\begin{align*}
  V_{1}(t+1) &= V_{1}(t) + SV(t) -  VV_{1}(t) - V\!E_{1}(t),\\
  V_{2}(t+1) &= V_{2}(t) + VV_{1}(t) - VV_{2}(t) - V\!E_{2}(t),\\[0.1cm]
  V_{3}(t+1) &= V_{3}(t) \\ &\,\,\,\,\,\,+ VV_{2}(t) - \mathrm{daily\_second}_{3,t}  - V\!E_{3}(t),\\[0.1cm]
  V_{4}(t+1) &= V_{4}(t) \\ &\,\,\,\,\,\,+ VV_{3}(t) - \mathrm{daily\_second}_{4,t} - V\!E_{4}(t),\\[0.1cm]
  V_{5}(t+1) &= V_{5}(t) \\& \!\!\!\!\!\!\!\!\!\!\!\!\!\!+ \mathrm{daily\_second}_{3,t}  + \mathrm{daily\_second}_{4,t} - V\!E_{5}(t),    \\[0.1cm]
  E_{V}(t+1)   &= E_{V}(t) + \textstyle{\sum_{j=1}^{5}}V\!E_{j}(t) - EI_{V}(t),   \\
  I_{V}(t+1)   &= I_{V}(t) + EI_{V}(t) - IR_{V}(t) - IU_{V}(t),\\
  U_{V}(t+1)   &= U_{V}(t) + IU_{V}(t) - U\!R_{V}(t),\\
  R_{V}(t+1) &= R_{V}(t) + IR_{V}(t) +U\!R_{V}(t).
\end{align*}
The values of $\psi_j$ from Table \ref{tab:efficacy}
are used to obtain the following expression for $p_{V,IU}$: 
%
%
\begin{align*}
\label{eq:icu_prob_v}
  p_{V,IU} &(= p_{V,IU,t}) 
  \\ &\quad = p_{IU}\times 
  \sum_{j=1}^{5} (1-\psi_{j})\,\tfrac{V_{j}(t)}{\sum_{k=1}^{5} V_{k}(t)},
\end{align*}
that is, amongst vaccinated infected individuals the probability of ICU admission is reduced based on a weighted average that involves the protection parameters $\psi_j$. Alternatively, one could consider separating $I_{V}(t)$ into $5$ compartments with separate rates but this would lead to the introduction of inessential complexities into the model.

\begin{remark}
In the applications we treat $p_{IU}$, $p_{U\!R}$ and $p_{VV}$ as known, given available results from the literature and concrete evidence for the performance of the vaccinations. We will also fix $p_{EI}$, $p_{I\!R}$ to values obtained using the data, as our main interest lies in the interaction parameter $\beta$. 
So, later on we will not include the above probabilities as components of the  model parameter vector $\theta$.
\end{remark}

\subsection{Data and Likelihood}
\label{subsec:like}
The model thus far described  defines the discrete-time Markov process $\{X_t\}$. 
That is, the dynamics defined in Section \ref{subsec:state} give rise to a transition kernel
%
$X_t\,|\,X_{t-1} \sim p(X_t\,|\,X_{t-1};\theta)$,
for the parameter vector
$\theta = (\beta, p_{EI}, p_{IR})$. 
We will treat $\{X_t\}$ as a latent process in the formulation of an HMM. As explained earlier, such a setting will allow accessing the machinery of computational SMC methods to learn the posterior of latent variables and parameters via powerful sequential algorithms.  

Our data, $\{Y_t\}$, correspond to reported daily ICU counts due to COVID-19 infections. Recall that the latent state $X_t$ contains the compartments $ICU(t)$ and $ICU_{V}(t)$ which also refer to ICU numbers.  
We write the sum of unvaccinated and vaccinated ICU compartments as:
\begin{equation}
  H(t) = ICU(t) + ICU_{V}(t).
\end{equation}
We model the observed reported ICU counts $\{Y_t\}$ via a Negative-Binomial distribution centred at $H(t)$, with overdispersion parameter 
$k_{\mathrm{obs}}$, following prior work that highlights the robustness of such a distribution in capturing overdispersed epidemic count data~\cite{cori2013framework, Lloyd-Smith2005}. That is, we have: 
\begin{equation*}
\label{eq:obs_model}
Y_t\,|\,X_t \sim \mathrm{NegBinomial}
\bigl(k_{\mathrm{obs}},\,p_t\bigr),
\end{equation*}
where we have set: 
\begin{equation*}
p_t = \frac{k_{\mathrm{obs}}}{k_{\mathrm{obs}} + H(t)}.
\end{equation*}
Note that we indeed have 
$\mathrm{E}\,[\,Y_t\,|\, X_t\,] = H(t)$. In our mumerical results we fix $k_{\mathrm{obs}}=10$.

We have now set up an HMM along the lines of Section \ref{sec:HMM}.

\subsection{Actions and Updated Model}
\label{subsec:action}

Within our MDP setting, we correspond `actions' to varying levels of NPIs. 
In terms of the effect on the model dynamics, we allow the infection rate parameter $\beta$ to assume a different value for each NPI level.   

In particular, we consider a set of integers:
\begin{equation*}
a_t\in\{1,\ldots, A\}, 
\end{equation*}
with $A\ge 1$, representing increasing order of  NPI stringency (and effectiveness). 
In what follows we use $A=4$, with NPI states corresponding to `no intervention', `mild closures' (e.g., the rule of six), `semi-lockdown' (e.g., working from home) and `full lockdown'.
The infection rate $\beta$ depends  heavily on such NPIs, so we extend our model from the one which utilises a single $\beta$ described in Section \ref{subsec:state} (see \eqref{eq:beta1}, \eqref{eq:beta2}) by introducing an ordered sequence of parameters:
\begin{equation*}
\beta_1 > \cdots > \beta_A.
\end{equation*}
The above modelling choice is inspired by the 
timeline of UK government NPIs from March 2020 to December 2021 \cite{IFG2022}.

Mathematically, the `force of infection' $\lambda_t$ \cite{keeling2008modeling} 
at time $t$ writes as: 

\begin{equation*}
\lambda_t
= \bigl(
\sum_{j=1}^{A}\beta_j\,\mathbf{1}_{\{a_t=j\}}\bigr)\,
\tfrac{I(t) + I_{V}(t)}{N}.
\end{equation*}



\subsection{Model Parameters}

Recall that the model parameters $\varepsilon_j$, $\psi_j$, $1\le j\le 5$, representing vaccine efficacies, 
are fixed at the values given in Table \ref{tab:efficacy}, based on relevant  information from previous studies. Similarly, we fix the following parameter values:  
\begin{itemize}
  \item[-] $p_{IU} = 1-e^{-1/10}$;
  \item[-] $p_{VV} = 1/20$: daily vaccine waning rate;
  \item[-] $p_{U\!R} = 1/10$: daily ICU discharge/recovery rate.
\end{itemize}
The above values are obtained from studies in the literature \citep{bernal2021effectiveness,sheikh2021sars,icnarc2020covid}.
For $p_{EI}$, $p_{IR}$, we obtain fixed values by, first, learning their posterior given $Y_{1:T}$ and then fine-tuning the estimated posterior medians by minimising a weighted root mean square error (RMSE) between model-simulated and observed ICU occupancy, using a subset of the UK ICU dataset that covers March--December 2020. The optimisation was carried out with the differential evolution global algorithm \cite{virtanen2020scipy}, which is well suited to noisy, non-convex objectives. This calibration ensures that the model realistically reproduces observed ICU dynamics. 
The values obtained are provided below:
\begin{itemize}
  \item[-] $p_{EI} = 1-e^{-1/9.86}$;
  \item[-] $p_{IR} = 1-e^{-1/10.41}$.
\end{itemize}
Thus, the parameter vector of interest involves the interaction rates and is defined as follows
\begin{equation}
\theta = \beta_{1:4}.
\label{eq:theta}
\end{equation}
\subsection{Real \& Simulated Data}
\label{subsec:counter}

Real data, denoted $Y_{1:T}$, correspond to publicly available ICU numbers, due to COVID-19 infection, between March 2020 and June 2021 \cite{NHSEnglandCOVIDHospitalActivity}. 

We also need simulated data for the `idealised' scenarios when the epidemic is assumed to evolve under the decision-making approach induced via the RL machinery. We generate such data via an `idealised' model corresponding to the one already described above  
obtained via the following procedure. We use SMC\textsuperscript{2} to learn the posterior distribution  $p(\theta|Y_{1:T})$, for $\theta=\beta_{1:4}$ and make appropriate use of samples obtained from such a posterior.  




\subsection{Reward Function}
\label{subsec:reward}

The choice of reward function is critical for formulating an MDP that attains intended objectives. In our epidemic setting
the reward function must balance two competing considerations: first, minimising the health burden of ICU occupancy; second, accounting for the socioeconomic costs of restrictive interventions.  
We define below a reward function derived through empirical simulations and informed discussions with domain experts. While the given formulation is chosen for its illustrative value and realistic trade-offs, the precise specification of the reward is not a core contribution of this work. Instead, the reward structure can be freely modified by policymakers or other users of this framework to reflect varying policy priorities or constraints. Indeed, the algorithmic framework we will later provide for solving the RL problem at hand can accommodate general reasonable specifications of the reward.
In general, the design of specific principled intervention cost functions is a challenging task due to the inherent difficulty of quantifying societal loss, preferences and trade-offs in public health decision-making.

We proceed as follows. For the positive constants $T_{\mathrm{crash}}$, $\kappa_{\mathrm{icu}}$, $\kappa_{\mathrm{so-ec}}$, we specify the following reward function $r_t = r(y_t,a_t)$: 
%
\begin{equation}
\label{eq:reward}
r_t = \left\{
\begin{array}{ll}
  -P,
    &\!\!\!\!\text{if}\,\,\, y_t>T_{\mathrm{crash}},\\[0.3cm]
  -\kappa_{\mathrm{icu}}\cdot y_t &  
    \\ \,\,\,\,\,\,\,\,\,
    -\kappa_{\mathrm{so-ec}}\;C(a_t,\ell_t),&\,\,\text{otherwise.}
\end{array}
\right.
\end{equation}
$C(a_t, \ell_t)$ is an `intervention cost function', 
with $\ell_t$ being a counter that tracks the number of consecutive days (up to, and including, $t$) the agent has maintained an intervention of the same severity as the current one at $t$ or a more strict one (if $a_t=1$ then $\ell_t=0$). 
The constants $P$, $\kappa_{\text{icu}}$, $\kappa_{\text{so--ec}}$ and $T_{\text{crash}}$ represent (respectively) a high cost, the penalty per occupied ICU places, the weight for intervention-related socioeconomic costs and the ICU capacity threshold beyond which `catastrophic' outcomes are assumed. 
These numbers can be adjusted to reflect different policy priorities or healthcare system constraints. For our numerical results we set $P=10^5$,  $T_{\mathrm{crash}}=6,000$.

The intervention cost function $C(a_t, \ell_t)$ assigns a penalty based on the action $a_t$ and the current length $\ell_t$. The cost structure could aim to capture the increasing social and economic burden of prolonged or intensified interventions, with different growth profiles for each level. We have made the following illustrative specificaton: 
\begin{equation}
\label{eq:cost}
C(a_t, \ell_t) = 
\begin{cases}
0, & a_t = 1, 
\\
50 \cdot \log(1 + \ell_t), & a_t = 2, 
\\
200 \cdot \ell_t, & a_t = 3, 
\\
800 \cdot \ell_t, & a_t = 4. 
\end{cases}
\end{equation}



Consideration of $\ell_t$  leads to a specification that captures both the immediate public health cost of high ICU load and the longer-term burden of sustained or escalating social restrictions. 




\section{Our MDP Formulation}
\label{sec:MDP}

We summarise the developments in Section \ref{sec:1} to describe the MDP setting at hand. We have set up the tuple $(\mathcal{S},\,\mathcal{A},\,\mathcal{P},\,r,\,\gamma)$ comprised of the following elements. The Markov chain $S_t = (X_t, Y_t)\in \mathcal{S}$ is assumed to be a Markovian process, with transition dynamics $\mathcal{P}$ modelled as in Sections \ref{subsec:state}--\ref{subsec:like}. 
The reward function $r(\cdot)$ is defined in Section \ref{subsec:reward}. Note, from (\ref{eq:cost}), that the specification of reward seems to go beyond a standard MDP setting, in the sense that the `length' $\ell_t$ looks at a number of past states of $\{S_t\}$. However, one can re-define a Markov chain which contains $\ell_t$ as a state, thus obtaining a standard MDP and overcoming this technicality. Such an approach for constructing  a Markovian system is also used in statistical models involving change points, see e.g.~\cite{yild:13}. For simplicity, we avoid such a reconstruction of our MDP system and retain the specification as already described, keeping the above point in mind.

\section{Real-Time RL}
\label{decision tool}

\subsection{Planning Structure}
Our framework implements a sequential decision-making process under uncertainty, aimed at dynamically optimising NPIs in real-time. At each decision point $t_0$, we use all data
available up to that day (collected as described in Section \ref{subsec:counter}) to update our beliefs about the `free' epidemic parameter vector $\theta$ in (\ref{eq:theta}) 
and the hidden state $X_{t_0}$ via Bayesian inference. For such a sequential task we make use of the 
SMC\textsuperscript{2}-algorithm \citep{ChopinJacobPapaspiliopoulos2013}. SMC\textsuperscript{2} takes advantage of the HMM-structure of our epidemic model to deliver samples from the sequence of posteriors for the parameters and the hidden state as we move forwards in time, in a computationally effective manner. Given the current posterior, we then perform a planning step, where we implement two approaches for the development of policies -- recall that the state vector of the MDP writes as $S_t=(X_t,Y_t)$: 
\begin{itemize}
\item[(\textrm{I})] A policy, $\pi_{\phi}^{\textrm{I}}\equiv\pi^{\textrm{I}}_{\phi}(y_t)$, defined as a deterministic function of the observed ICU numbers, with the parameter vector $\phi$ corresponding to thresholds that trigger alternative actions. For this direction, $\phi$ will be learned via
roll-out of several simulated trajectories whilst taking under account the uncertainty for $\theta$ given the observations $y_{1:t}$ seen up to the present time. 
\item[(\textrm{II})] A policy $\pi^{\textrm{II}}\equiv \pi^{\textrm{II}}(y_t)$ of no specific restriction in its structure (in contrast to $\pi^{\textrm{I}}$ above). Here, we exploit the relatively small size of state (once $Y_t$ is partitioned into a number of groups) and action spaces and implement the $Q$-learning algorithm, again taking under consideration the uncertainty about $\theta$.  
\end{itemize}
%

Once the optimal action (i.e.~the NPI to implement) is chosen under the policy approach to be followed, we deploy it in the simulated `counterfactual' world, thus replacing the real  government intervention for that period. 
As mentioned earlier, since such decisions may differ from the ones taken by the government, the resulting epidemic trajectories will be simulated as described in Section~\ref{subsec:counter}. 

\begin{figure*}[!t]
  \centering
  \includegraphics[width=0.95\textwidth]{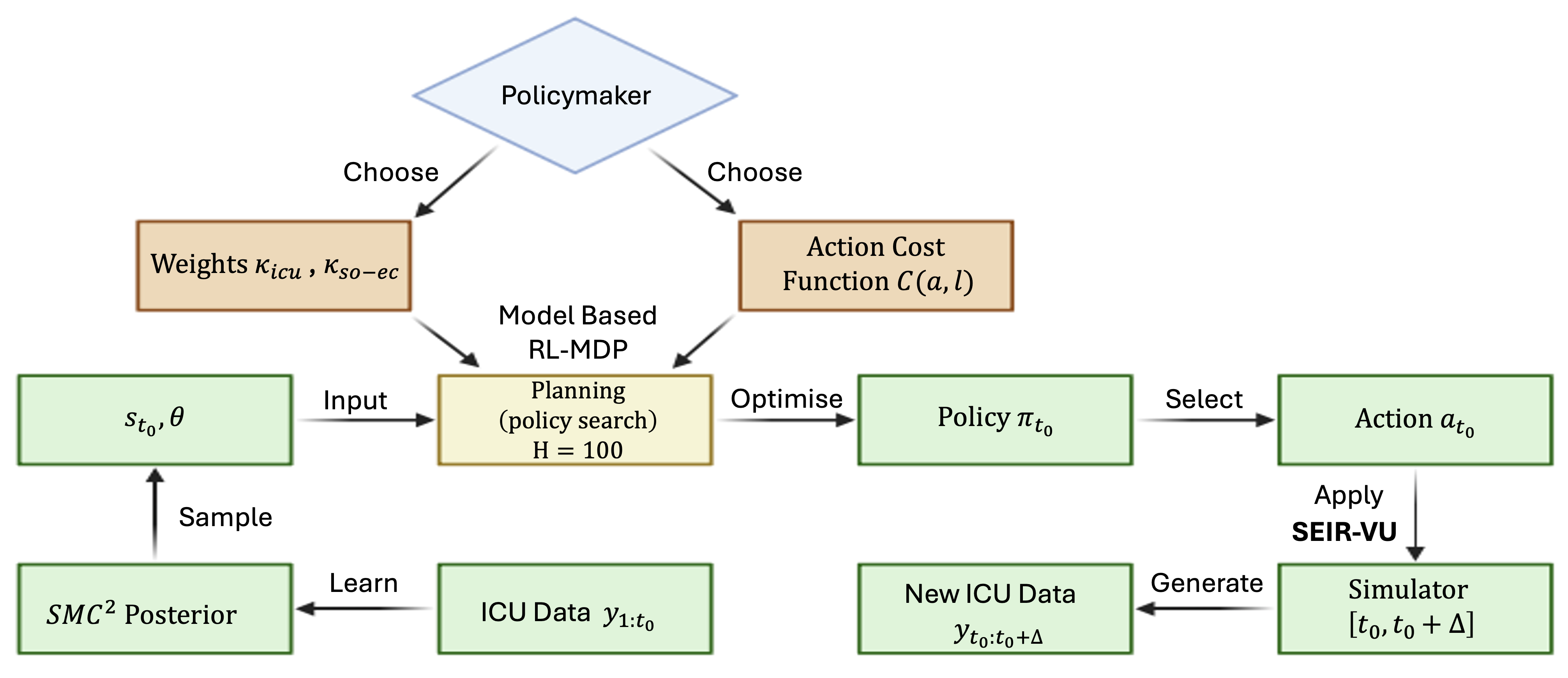}
  \caption{One-block decision loop. At decision-time $t_0$, ICU data update the SMC\textsuperscript{2} posterior; state and parameter $(s_{t_0},\theta)$ feed into a planner (horizon $H = 100$) yielding policy 
  ${\pi}_{t_0}$ and action $a_{t_0}$; the model generator runs to $t_0 + \Delta$, producing `counterfactual' ICU data for the next update.}
  \label{fig:one_block_loop}
\end{figure*}

To reproduce realistic scenarios, an update of the action is attempted every 
$\Delta$ days (for a decision period input $\Delta>0$): the agent collects new observations (on the ICU occupancy), updates the posterior over the model parameter vector $\theta$ and the latent state, plans the next intervention based on the updated beliefs, and deploys a new action. 
The described decision-support framework is illustrated in Figure~\ref{fig:one_block_loop}.

Recall that we have access to ICU data for the period 
between March 2020 and June 2021 \cite{NHSEnglandCOVIDHospitalActivity}, and that the agent has no knowledge of the true transmission rates. 
Initiating the decision period in March 2020 with a uniform prior over the 
$\beta_j$'s would lead to unreliable early decisions.
Thus, we instead consider a 300-day decision period that starts in August 2020 and we use the historical data during 
March--August 2020 to obtain a posterior (in the sense of acquiring a Monte Carlo sample representation) of the model vector parameter $\theta$ in (\ref{eq:theta}). Such a posterior then serves as the initial distribution when the decision-making horizon begins. In this manner, our framework mimics a policymaker who enters the decision period with a realistic prior knowledge calibrated upon earlier epidemic information.

Thus, in our setting, the decision horizon is 
$T_{\mathrm{HORIZON}}=300$ days, and is subdivided into
$B = T_{\mathrm{HORIZON}}/\Delta = 30$ consecutive decision blocks, each of length $\Delta = 10$ days. At the start of each block $b$ (corresponding to day $t_{0}=(b-1)\,\Delta$), $b=1,\ldots, B$, 
we apply RL under either of the two policies under consideration, $\pi_{\phi}^{\textrm{I}}$ and $\pi^{\textrm{II}}$. 
In particular, $\pi_{\phi}^{\textrm{I}}$ is obtained via consideration of the maximisation (we set $\mathcal{Y}_{t_0}:=\{y_{0:t_0}\}$): 
\begin{align*}
\arg \max_{\phi} \mathrm{E}_{\theta,x_{t_0}|\mathcal{Y}_{t_0}} 
\Big[\,\sum_{t=t_0}^{t_0+H}\gamma^{t-t_0} r\big(y_t,\pi_{\phi}^{\mathrm{I}}(y_t)\big)\,\Big],
\end{align*}
for a time horizon $H=100$ chosen to allow enough steps forwards, whilst also taking under consideration the relatively small $T_{\textrm{HORIZON}}$ of the problem at hand.
In practice, the above expectation is approximated via a Monte Carlo procedure and the optimisation is carried our via a grid search. 
That is, we use the output of SMC\textsuperscript{2} to obtain $K=25$ samples $\{(\theta^{(k)},x_{t_0}^{(k)})\}_{k=1}^{K}$ from the current posterior  $p(\theta,x_{t_0}|\mathcal{Y}_{t_0})$, and use these samples to generate corresponding Monte Carlo roll-outs of the stochastic SEIR--VU model.
In the case of the second policy $\pi^{\textrm{II}}$, a standard $Q$-learning iteration is applied over the same time horizon $H=100$. Full details are provided in the sequel.



\subsection{Threshold-Based Policy $\pi^{\textrm{I}}$}
\label{Treshold}
In each decision block $b$ of $\Delta$ days, started at day $t_0=(b-1)\Delta$, $b=1,\ldots, B$, we utilise $K$ parameter and latent state samples provided by SMC\textsuperscript{2}:
\begin{equation*}
(\theta^{(1)},x_{t_0}^{(1)}),\ldots,(\theta^{(K)},x_{t_0}^{(K)})\sim p(\theta,x_{t_0}|\mathcal{Y}_{t_0}).
\end{equation*}
We thus make use of the current posterior distribution to capture the uncertainty in the epidemic dynamics. We then consider a rich family of deterministic policies defined by placing thresholds on the ICU occupancy, indexed by the triples:
\begin{align}
\label{eq:phi}
\phi=\{(\tau_{1},\tau_{2},\tau_3):0<\tau_{1}<\tau_{2}<\tau_{3}\}.
\end{align}
%
Each $\phi$-vector induces a simple rule that maps the observed ICU load to one of four intervention intensities, offering  a transparent and interpretable control, in particular: 
\begin{equation}
\label{eq:threshold_policy}
\pi_{\phi}^{\textrm{I}}(y_t) =
\begin{cases}
1, & y_t < \tau_{1},\\
2, & \tau_{1} \le y_t < \tau_{2},\\
3, & \tau_{2} \le y_t < \tau_{3},\\
4, & y_t \ge \tau_{3}.
\end{cases}
\end{equation}
To apply the corresponding finite-horizon RL approach and find the optimal policy, we generate Monte Carlo roll-outs of the stochastic SEIR--VU model under each $(\theta^{(k)},x_{t_0}^{(k)})$ over a look-ahead period of $H=100$ days, accumulate daily costs that trade-off ICU burden against intervention severity and average these totals across the $K$ simulations to obtain an expected cost estimate $\hat{C}_{\phi}$. That is, one first calculates:
\begin{equation}
\label{eq:rollout_cost}
C_{\phi}(\theta^{(k)},x_{t_0}^{(k)})
= \sum_{t=t_0}^{t_0+H} \gamma^{t-t_0} r(y_t^{(k)}, \pi^{\textrm{I}}_{\phi}(y_t^{(k)})),
\end{equation}
where $r(\cdot, \cdot)$ is defined in (\ref{eq:reward}), and $y_{t_0}^{(k)}$ coincides with the `real' observation $y_{t_0}$. We then obtain the Monte Carlo estimate:
\begin{equation}
\label{eq:avg_cost}
\hat{C}_{\phi}
= \tfrac{1}{K}\sum_{k=1}^{K}C_{\phi}(\theta^{(k)},x_{t_0}^{(k)}).
\end{equation}
%
The estimated mean cost 
$\hat{C}_\phi$ is evaluated across a number of grid points for $\phi$ so that we can identify the threshold triple $\phi^{*}$ that minimises $\hat{C}_\phi$. We then deploy the identified optimal policy $\pi^{\textrm{I},*}=\pi^{\textrm{I}}_{\phi^*}$ for the next
$\Delta$ days, collect the next set of ICU observations, update the posterior to $p(\theta,x_{t_0+\Delta} |\mathcal{Y}_{t_0+\Delta})$ and obtain samples from the latter via SMC\textsuperscript{2}.
This procedure is iterated until we reach time $T_{\textrm{HORIZON}}$.
Such a sequential re‐planning loop combines the computational efficiency and interpretability of threshold rules with the adaptivity of Bayesian updating and look-ahead Monte Carlo simulation, thus approximating finite-horizon MDP planning without explicit dynamic programming.

The pseudocode in Algorithm~\ref{alg:smc2_gridsearch} summarises all steps of the above sequential decision approach with interpretable planning.

\begin{algorithm*}[!t]
\small  
\caption{Finite-Horizon MDP for ICU-Thresholded Policy $\pi_{\phi}^{\textrm{I}}$}
\label{alg:smc2_gridsearch}
\KwIn{%
 $T_{\textrm{HORIZON}}$; ICU data 
 $\{y_t\}_{0:T_{\textrm{HORIZON}}}$; 
 vaccination streams $\{\mathrm{daily\_first}_t,\mathrm{daily\_second}_t\}_{0:T_{\textrm{HORIZON}}}$;\\
  decision period $\Delta$; training horizon $H$; posterior draws $K$; particles $N$;\\
  crash limit $T_{\textrm{crash}}$, cost $P$; weights $(\kappa_{\textrm{icu}},\kappa_{\textrm{so-ec}})$; discount $\gamma$; Grid $\texttt{GRID}$.
}
\BlankLine
\KwOut{%
  Rewards $r_{0:T_{\textrm{HORIZON}}}$; ICU trajectory $y_{0:T_{\textrm{HORIZON}}}$; actions $a_{0:T_{\textrm{HORIZON}}}$. 
}
\BlankLine
%
%
%
%
\BlankLine
\For{
{$b=1$ \KwTo $B\,(\,=T_{\mathrm{HORIZON}}/\Delta\,)$}}{ \vspace{0.1cm}
  Set $t_0 = (b-1)\Delta$;
  \BlankLine
  \For{{$\phi \in \emph{\texttt{GRID}}$}}{ 
  \BlankLine
  \For{$k=1$ \KwTo $K$}{
    Sample $(\theta^{(k)},x_{t_0}^{(k)})\sim p(\theta,x_{t_0}|\mathcal{Y}_{t_0})$\;
    Calculate $C_{\phi}(\theta^{(k)},x_{t_0}^{(k)})
= \sum_{t=t_0}^{t_0+H} \gamma^{t-t_0} r(y_t^{(k)}, \pi^{\textrm{I}}_{\phi}(y_t^{(k)}))$\;
  }
  Calculate $\hat{C}_{\phi}
= \tfrac{1}{K}\sum_{k=1}^{K}C_{\phi}(\theta^{(k)},x_{t_0}^{(k)})$\;
  }
  \vspace{0.1cm}
  Find $\phi^*=\arg\max_{\phi\in \texttt{GRID}}\hat{C}_{\phi}$\;
  \BlankLine
  \For{{$t=t_0$ \KwTo  $t_0+\Delta-1$}}{ 
  \vspace{0.1cm}
  Receive observation $y_t$\; 
  \vspace{0.05cm}
  Apply action $a_t = \pi^{\textrm{I}}_{\phi^*}(y_t)$\;
  \vspace{0.1cm}
  Calculate reward $r_t = r(y_t,a_t)$\;
  }
\BlankLine
 %
}
\end{algorithm*}

\subsection{Tabular-Based Policy  $\pi^{\textrm{II}}$}
\label{subsec:tabular}

In addition to the threshold‐based approach, we also consider  
a policy mapping a partition of the state space of $y_t$ to actions.
The corresponding optimal policy will in this case be approximated by  
a tabular $Q$‐learning scheme that respects parameter uncertainty via use of posterior sampling. 

\medskip
\textbf{Policy Specification.}
The (integer-valued) ICU loads $y_t$ at each time $t$ are mapped onto the finite number of states $\{1,2,\ldots, G\}$, for $G\ge 1$, via consideration of threshold positions, denoted as  $\textrm{THR}_{1:(G-1)}$, with $\textrm{THR}_1<\cdots <
\textrm{THR}_{G-1}$, so that state $g_{t}$ at time $t$ is obtained as follows:   
\begin{equation}
\label{eq:stateB}
g_{t} = \sum_{g=1}^{G} g\cdot \mathrm{I}\,[\,\textrm{THR}_{g-1}\le y_t\le \textrm{THR}_{g}\,],
\end{equation}
under the conventions $\textrm{THR}_{0}=0$ and $\textrm{THR}_{G}=\infty$. 
Similarly to the threshold-based policy, we allow for actions $a_t\in \{1,\ldots, A\}$, as defined in 
Section~\ref{subsec:action}. 
%

\medskip
\textbf{Policy Learning.}  
We consider the same blocking of decision times as in Section \ref{Treshold}.
Given a current decision time $t_0 = 
(b-1)\Delta$, the policy learning method proceeds as follows. 
We group trajectories into $\Delta$-day `slices' and perform a temporally aggregated $Q$‐learning update at the end of each slice. The learning horizon is denoted by $H$ and we assume (for simplicity) that $M:=H/\Delta$ is a positive integer. The estimation of the state-action value function involves $M$ iterates for each episode, and the iterates will be repeated for a number of episodes $E\ge 1$. The estimate of the state-action value function used at the start of each episode is the one obtained at the end of the immediately previous episode. For the very first episode, the initial estimate is the one obtained by applying the iterates provided below for the real data up-to the first decision time, starting from a trivial estimate (i.e.~constant values for all states and actions) at time $1$.  

Recall we are at decision time $t_0 =  (b-1)\Delta$. 
We collect $K$ samples $\{\theta^{(k)},x_{t_0}^{(k)}\}_{k=1}^K$ from the current Monte Carlo approximation of the  posterior distribution $p(\theta,x_{t_0}|\mathcal{Y}_{t_0})$ provided by SMC\textsuperscript{2}. 
For each $k$ the environment dynamics are determined by the SEIR--VU model with parameter $\theta=\theta^{(k)}$ and initial latent state value $x_{t_0}=x_{t_0}^{(k)}$. 
For a given $k$, we describe the learning process for one episode given a corresponding initial estimate $Q_0^{(k)}(g,a)$.
In particular, we simulate the SEIR--VU model for a look‐ahead time horizon of $H=100$ days, starting from state $x_{t_0}^{(k)}$ and an action $a_{t_0}^{(k)}$ determined as $\arg\max_{a} Q_0^{(k)}(g_{t_0}^{(k)},a)$, or via consideration of a random action (with small probability) in the context of a greedy $Q$-learning implementation. Let 
$R_{m,\rm slice}^{(k)}$ be the sum of daily rewards over slice $m$,  i.e.:
\begin{align}
\label{eq:R_slice}
R_{m,\mathrm{slice}}^{(k)} =\sum_{t=t_0+(m-1)\Delta}^{t_0+m\Delta-1} 
r(y_t^{(k)},a_{t_0+(m-1)\Delta}^{(k)}).
\end{align}
%
Notice that the action remains the same over the period of the slice, as such a consideration matches the real decision-making conditions.
We set:
\begin{gather*}
g=g_{t_0+(m-1)\Delta}^{(k)}, \quad a=a_{t_0+(m-1)\Delta}^{(k)}, \\ g'=g^{(k)}_{t_0+m\Delta},
\end{gather*}
i.e.~state $g'$ denotes the ICU-bin at the beginning of slice~$m+1$. 
The $Q$-learning update for such visited $(g,a,g')$ writes as:
\nopagebreak[4]
\begin{equation}
\label{eq:ql_update}
\begin{aligned}
Q_{m+1}^{(k)}(g,a)
&= Q_{m}^{(k)}(g,a) \\
&\quad + \alpha_{m}\Bigl(
   R_{m,\mathrm{slice}}^{(k)}
   + \gamma\,\max_{a'} Q_{m}^{(k)}(g',a') \\
&\qquad\qquad
   - Q_{m}^{(k)}(g,a)
\Bigr).
\end{aligned}
\end{equation}

Here $\gamma$ is interpreted as a discount factor with the time unit being $\Delta$ days instead of a single day.
Repeating the above update for $M$ slices, and a number of episodes $E$, we obtain the $Q$-table corresponding to the very last iteration, say $Q^{(k)}_{E,M+1}(g,a)$, over all $g\in \mathcal{G}$, $a\in\mathcal{A}$.  The estimates over $k$ are then averaged pointwise to yield a Bayes‐averaged state‐action value function:
\begin{equation}
\label{eq:q_average}
\bar{Q}(g,a)
= \tfrac{1}{K}\sum_{k=1}^{K}Q_{E,M+1}^{(k)}(g,a),
\end{equation}

\medskip
\textbf{Action Decision.}
The block‐level decision at time  $t_0 =  (b-1)\Delta$ is simply:
\begin{equation}
\label{eq:opt_action_q}
a_{b}^{\star}
= \arg\max_{a}\bar{Q}(g_{t_0},a),
\end{equation}
and is executed for the next $\Delta$ days in the real environment. 
Here $g_{t_0}$ denotes the discretised ICU‐load real state at the start of block 
$b$ as explained in (\ref{eq:stateB}). Observed ICU data over that interval are used to update the posterior via SMC\(^2\), and the procedure repeats for block $b+1$. Over the full time period of interest from $t=0$ to $t=T_{\textrm{HORIZON}}$, the receding‐horizon algorithm produces a sequence of Bayes‐optimal block‐level actions:
\begin{align*}
(a_{1}^{\star},\dots,a_{B}^{\star}).
\end{align*}

\medskip
\textbf{Summary.}
The above-described algorithm thus implements a receding‐horizon, Bayes‐risk‐aware control strategy by iteratively sampling from the posterior, performing slice‐aggregated, model-based hybrid $Q$‐learning under each simulated environment, averaging the resulting state‐action value functions, and deploying the greedy policy for each decision block.
The detailed implementation of this SMC\textsuperscript{2}-guided, posterior‐averaged $Q$-learning procedure is presented in Algorithm~\ref{alg:smc2_post_avg_q_learning} which summarises every step from posterior sampling and slice‐level updates to action deployment and posterior refresh.

\setcounter{algocf}{1}
\LinesNumbered
\setcounter{AlgoLine}{0}
\begin{algorithm*}[t!]
\small
\caption{Posterior-Averaged $Q$-Learning}
\label{alg:smc2_post_avg_q_learning}
\KwIn{
$T_{\textrm{HORIZON}}$; ICU data 
 $\{y_t\}_{0:T_{\textrm{HORIZON}}}$; 
 vaccination streams $\{\mathrm{daily\_first}_t,\mathrm{daily\_second}_t\}_{0:T_{\textrm{HORIZON}}}$;\\
  decision period $\Delta$; training horizon $H$; no of episodes $E$; posterior draws $K$; particles $N$;\\
  crash limit $T_{\textrm{crash}}$, cost $P$; weights $(\kappa_{\textrm{icu}},\kappa_{\textrm{so-ec}})$; discount $\gamma$; \\
      no of bins $G$, bin borders $\mathrm{THR}_{1:(G-1)}$; initial $Q$-table $Q_{\mathrm{start}}$.\\ 
    }
    \BlankLine
\KwOut{$r_{0:T_{\textrm{HORIZON}}}$; ICU trajectory $y_{0:T_{\textrm{HORIZON}}}$; actions $a_{0:T_{\textrm{HORIZON}}}$}
\BlankLine
\BlankLine
\For{$b=1$ \KwTo $B \,(\,=T_{\mathrm{HORIZON}}/\Delta\,)$}{ \vspace{0.1cm} 
   Set $t_0=(b-1)\Delta$\; 
   Collect $y_{t_0}$; Obtain corresponding $g_{t_0}$\; 
   \vspace{0.1cm} 

  \For{$k=1$ \KwTo $K$}{ \vspace{0.1cm}
    Sample $(\theta^{(k)},x_{t_0}^{(k)})\sim p(\theta,x_{t_0}|\mathcal{Y}_{t_0})$\; 
    Initialise $Q_{0,M+1}(\cdot, \cdot) = Q_{\mathrm{start}}(\cdot, \cdot)$\;\vspace{0.1cm}
   \For{$e=1$ \KwTo $E$}{ \vspace{0.2cm}
   Initialise $Q_{e,1}(\cdot, \cdot) = Q_{e-1,M+1}(\cdot, \cdot)$\;\vspace{0.2cm}
   \For{$m=1$ \KwTo $M\,(\,=H/\Delta\,)$}{ \vspace{0.2cm}
    Determine action $a_{e,t_0+(m-1)\Delta}^{(k)}$ via greedy $Q$-learning based on 
    current $Q^{(k)}_{e,m}$\;\vspace{0.1cm}
    Calculate $R_{e,m,\mathrm{slice}}^{(k)} =\sum_{t=t_0+(m-1)\Delta}^{t_0+m\Delta-1} 
    r(y_{e,t}^{(k)},a_{e,t_0+(m-1)\Delta}^{(k)})$\; \vspace{0.1cm}
    Set $g=g_{e,t_0+(m-1)\Delta}^{(k)}$, $a=a_{e,t_0+(m-1)\Delta}^{(k)}$, $g'=g^{(k)}_{e,t_0+m\Delta}$\; \vspace{0.1cm}
    Update $Q_{e,m+1}^{(k)}(g,a) = Q_{e,m}^{(k)}(g,a) 
                 + \alpha_{m}\big(
                   R_{e,m,\mathrm{slice}}^{(k)}
                   + \gamma\,\max_{a'}Q_{e,m}^{(k)}(g',a')
                   - Q_{e,m}^{(k)}(g,a)\big)$\;
}
  }
  }
  \vspace{0.1cm}
  Calculate $\bar{Q}(g,a)
= \tfrac{1}{K}\sum_{k=1}^{K}Q_{E,M+1}^{(k)}(g,a)$\;
 Select action $a_{b}^{\star}
= \arg\max_{a}\bar{Q}(g_{t_0},a)$\;
    \BlankLine
  \For{{$t=t_0$ \KwTo  $t_0+\Delta-1$}}{ 
  \vspace{0.1cm}
  Receive observation $y_t$\;
  \vspace{0.05cm}
  Apply action $a^{*}_b$\;
  \vspace{0.1cm}
  Calculate reward $r_t = r(y_t,a^{*}_b)$\;
  }
\BlankLine
 %
}
\end{algorithm*}

\section{Experimental Results}\label{results}

We present results on the empirical performance of the two sequential decision‐making algorithms introduced in the previous sections.
All experiments span the same 300-day horizon and make use of multiple replicates of counterfactual data to quantify such a source of uncertainty.  We compare each method against a random‐policy baseline and the real historical decisions, and present results over the generated 
NPI trajectories and the corresponding cost functions. 

\subsection{Model Validation}

To verify that our SEIR--VU model faithfully reproduces real‐world ICU dynamics, we replayed the actual sequence of UK government NPIs applied between August 2020 and June 2021.  At each day $t$, we supplied our model with the real action $a_t$ and simulated the evolution of the full state $S_t=(X_t,Y_t)$
under the true vaccination roll-out.
Each simulation used a sample from the posterior distribution of the parameter vector $\theta$ given the full data, $\mathcal{Y}_{T_{\textrm{HORIZON}}}$, between  August 2020 and June 2021.
We then compared the simulated trajectories of ICU numbers against the observed NHS England ICU data over the 300-day period.

\begin{figure*}[!ht]
  \centering
  \includegraphics[width=0.85\textwidth]{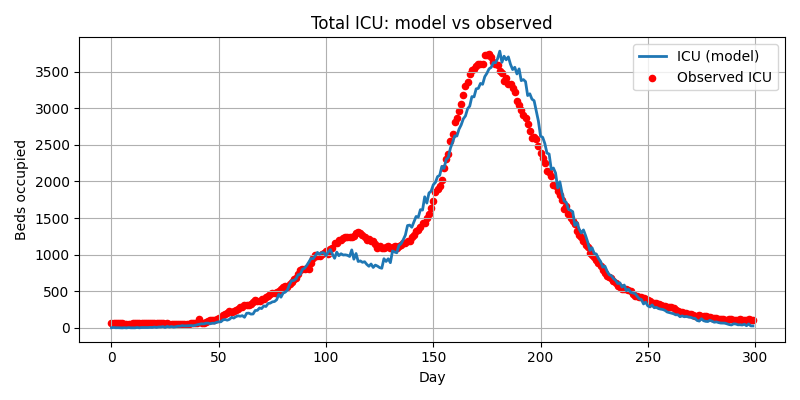}
  \caption{Observed ICU occupancy versus average values of simulated roll-outs obtained via the SEIR–VU model, under the real government NPIs.}
  \label{fig:validation}
\end{figure*}

Figure~\ref{fig:validation} shows in blue the average curve obtained via different samples from the posterior of $\theta$ and corresponding varying realisations of the stochastic model dynamics. It is apparent that the obtained curve follows closely the timing and magnitude of the actual ICU admissions (red dots), capturing both the peaks and troughs of the epidemic waves.  This close agreement indicates that our stochastic 
SEIR--VU model provides a reliable simulator for downstream decision‐making and policy evaluation.

\subsection{Data and Experimental Setup}

For all our experiments in the sequel we run $R=10$ independent replicates 
over the generation of the counterfactual data.  
We use a fixed  ICU weight $\kappa_{\mathrm{icu}}=1$ and three intervention weights $\kappa_{\mathrm{so-ec}}\in\{0.2,0.5,0.8\}$.  The decision horizon is $T_{\mathrm{HORIZON}}=300$ days, split into $B=30$ blocks of $\Delta=10$ days each. 
The value $\Delta=10$ is chosen to resemble the number of times of a new action was taken by the authorities during the pandemic. In particular, there were $20$ changes of policy, and our choice of $\Delta$ gives a maximum of $30$ policy changes.
In all cases, we set $\gamma=0.95$. For both policies we choose a crash threshold $T_{\mathrm{crash}}=5,000$ beds with a penalty of $P=-10^{6}$ and obtain posterior updates via use of SMC\textsuperscript{2} with $500$ particles.

For the case of policy $\pi^{\mathrm{I}}$, we 
provide the following details.
The optimisation over a grid of $\phi$  begins with a full evaluation over all $\binom{30}{3}=2300$ (ordered) triples for the threshold vector $\phi=(\tau_1,\tau_2,\tau_3)$ in (\ref{eq:phi}) obtained from a geometric grid of $30$ points in $[10,8000]$ along each of the $3$ thresholds. The objective function at decision time $t_0=(b-1)\Delta$
is estimated as in (\ref{eq:rollout_cost}) using $K=25$ posterior samples.
After the initial global search identifies a high‐performing threshold triple, we restrict subsequent searches to a neighborhood around the previous optimum rather than re‐exploring the entire grid.  This local‐refinement strategy is justified because the underlying epidemiological dynamics and posterior uncertainty evolve smoothly over successive $\Delta$-day blocks, so the Bayes-optimal ICU thresholds at block $b+1$ are unlikely to differ drastically from those at block $b$.  By searching only within a small margin from the last chosen 
$\phi$, we dramatically reduce the number of candidates, thereby cutting computational cost and also reducing the effect of spurious fluctuations due to Monte Carlo noise. 

For for the tabular-based policy $\pi^{\textrm{II}}$, we partition the values of $y_t$ into $G=200$ bins, using a geometric grid on $[1,6000]$. Then,   
the policy learning procedure draws $K=25$ parameter vectors at each decision time, runs $E=80,000$ slice‐level episodes over a planning horizon of $H=100$ days, employs an $\varepsilon$–greedy exploration schedule decaying from $\varepsilon_{0}=0.20$ to $0.05$, and carries out updates via learning rates $\alpha_{k} = C/(C+k)$ with $C=45$.

\subsection{Model-Free $Q$-Learning}
\label{sec:naive_q_fails}

\begin{figure*}[t]
  \centering
  \includegraphics[width=\linewidth]{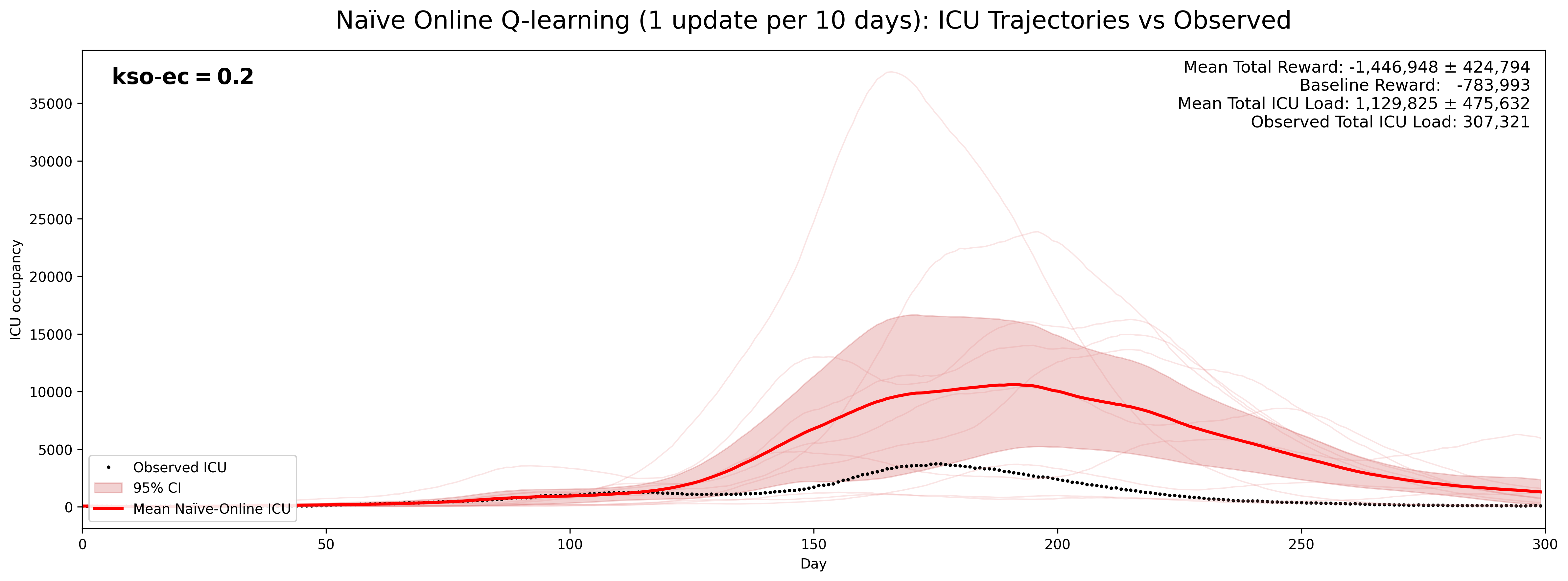}
  \caption{{Naïve online $Q$-learning (one TD update per $\Delta{=}10$ days) vs observed ICU.}
  }
  \label{fig:naive_online_bad}
\end{figure*}

Before adopting the described model-based $Q$-learning approach for policy $\pi^{\textrm{II}}$,  we investigate whether a `standard' \emph{model-free} $Q$-learning controller could provide reliable results in our setting.
Thus, we implement the canonical online $Q$-learning baseline \citep{SuttonBarto2018, WatkinsDayan1992}. I.e., at each decision time $t_0=(b-1)\Delta$, the controller selects an action via an $\varepsilon$-greedy decision-making w.r.t.~the current table $Q=Q(g,a)$, deploys the chosen action for $\Delta$ days, aggregates the realised block reward and performs \emph{one} update to $Q$ (one temporal-difference, `TD', update) before moving to the next block. Note that the $Q$-table is only updated at $B(=30)$ instances.

Figure~\ref{fig:naive_online_bad} displays the resulting trajectories. The policy reacts late and weakly: the mean path (recall that we use replicates over different realisations of counterfactual data) overshoots the observed peak and decays slowly, with wide uncertainty, leading to both considerably higher cumulative ICU load and lower reward against the true scenario.

We stress that $B=30$ TD updates are far too few for a tabular function $Q=Q(g,a)$ defined over $G=200$ ICU bins and $A=4$ actions; most $(g,a)$ pairs are rarely or never updated, so value information cannot propagate. Classical convergence of tabular $Q$-learning is asymptotic and requires infinitely many visits to each $(g,a)$ \cite{WatkinsDayan1992,BertsekasTsitsiklis1996,SuttonBarto2018}.
In our public-health setting (short horizon, few safe interventions, evolving dynamics) the classic online, model-free approach does not learn a useful value function surface in time. 
This is the key reason of opting for the \emph{model-based} alternative  described in Section \ref{subsec:tabular} and shown in detail in Algorithm \ref{alg:smc2_post_avg_q_learning}, which utilises: SMC$^{2}$-based sampling of parameters from the sequence of posteriors as more data become available; many simulated ($E>> 1$) planning episodes per block with horizon $H>\Delta$; posterior averaging of the obtained $Q$’s.
This approach provides sufficient updates and reduces variability, producing the stable block-wise convergence for the tabular function $Q$ reported later in Section~\ref{sec:convergence_q}.

\subsection{Policy Comparison}

To assess the relative merits of our two planners, we applied Algorithms~\ref{alg:smc2_gridsearch}--\ref{alg:smc2_post_avg_q_learning} for the same time horizon, alongside a random‐uniform policy and the actual historical NPI trajectory. 

\begin{figure*}[p]
  \centering
  \includegraphics[width=0.85\textwidth]{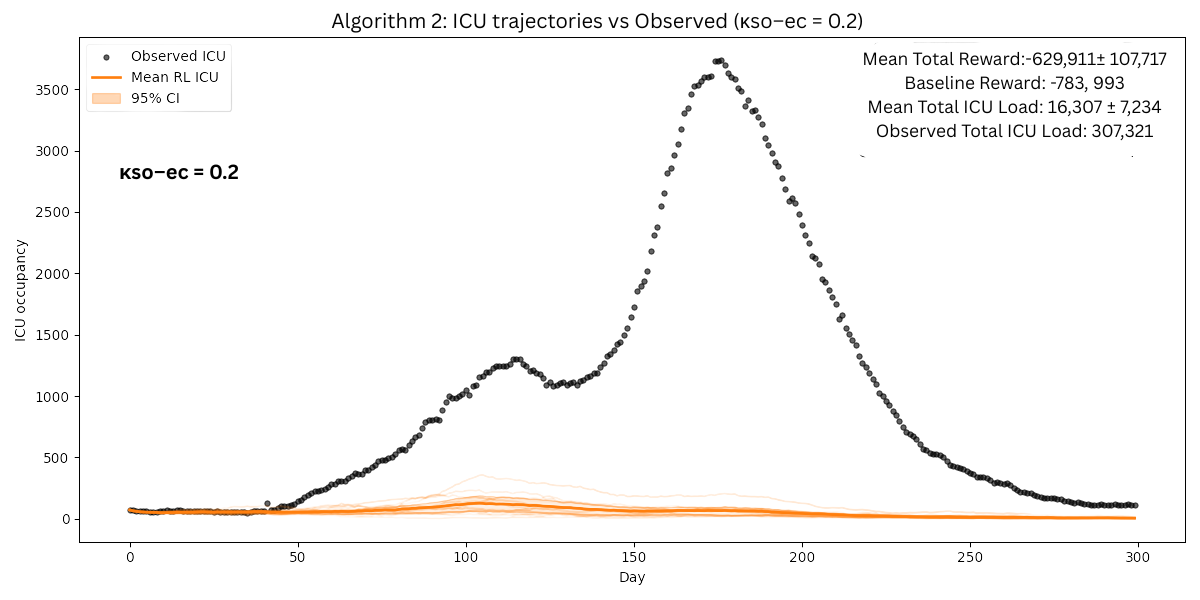}\\[1ex]
  \includegraphics[width=0.85\textwidth]{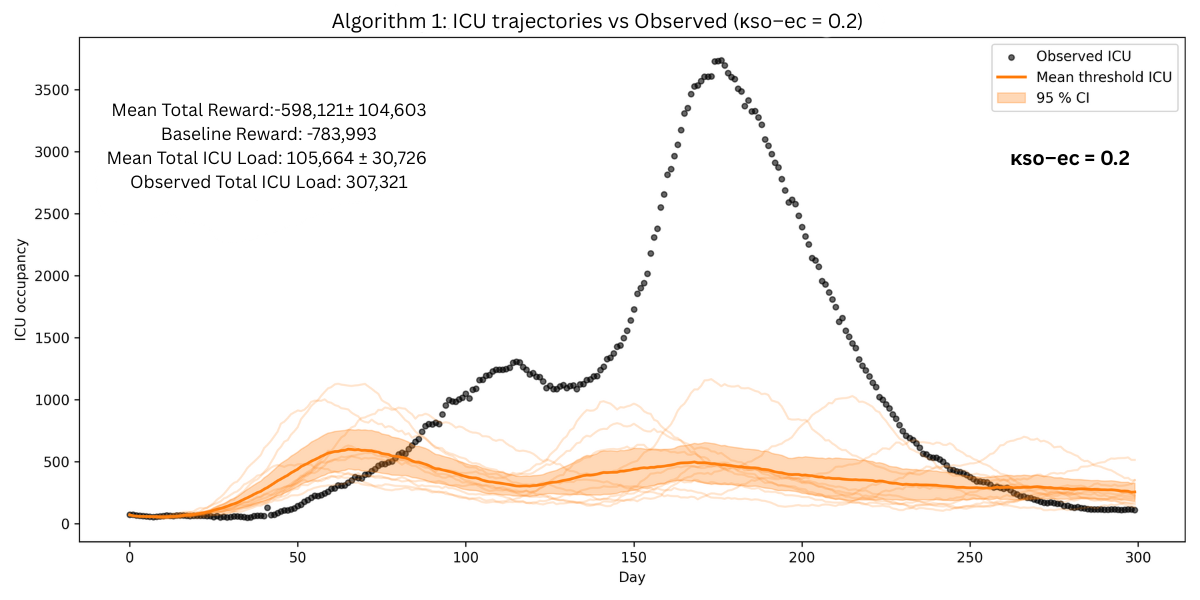}\\[1ex]
  \includegraphics[width=0.85\textwidth]{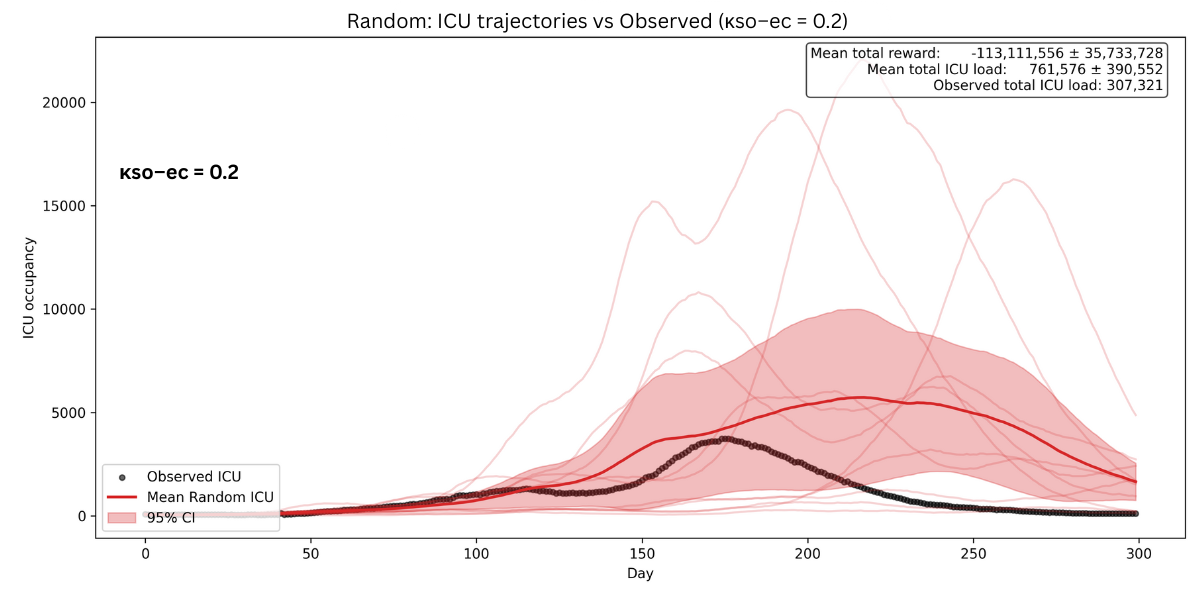}
  \caption{%
    Top Panel: ICU load under the $Q$‐learning controller with \(\kappa_{\textrm{so--ec}}=0.2\); 
    Middle Panel: ICU load under the threshold‐based controller with \(\kappa_{\textrm{so--ec}}=0.2\); 
    Bottom Panel: ICU load under a random policy with \(\kappa_{\textrm{so--ec}}=0.2\).
  }
  \label{fig:all_policy_comparison}
\end{figure*}

We conducted experiments with three different intervention cost weights, $\kappa_{\textrm{so--ec}}\in\{0.2, 0.5, 0.8\}$. As seen in Figure~\ref{fig:all_policy_comparison}, when $\kappa_{\textrm{so-ec}}=0.2$, Algorithms \ref{alg:smc2_gridsearch}--\ref{alg:smc2_post_avg_q_learning} outperform the real government implementation in terms of both ICU burden and cumulative reward for our given intervention cost (`Baseline Reward' indicates the total cost using the true governmental NPIs). With $\kappa_{\textrm{so--ec}} = 0.2$ (the smallest amongst the three values under consideration), the algorithms prioritise minimising the ICU burden versus reducing the intervention costs. The policies are thus prompted to apply stricter interventions and adopt a more aggressive approach to epidemic control, leading to a significant reduction in ICU burden. The random policy, also illustrated in Figure \ref{fig:all_policy_comparison}, due to its lack of strategic planning, fails to reduce ICU load effectively and has a very low reward. 
The improvements provided by the intelligent non-random policies are not due to chance, but the result of a well-structured decision-making framework.

\begin{figure*}[!ht]
  \centering
  \includegraphics[width=0.85\textwidth]{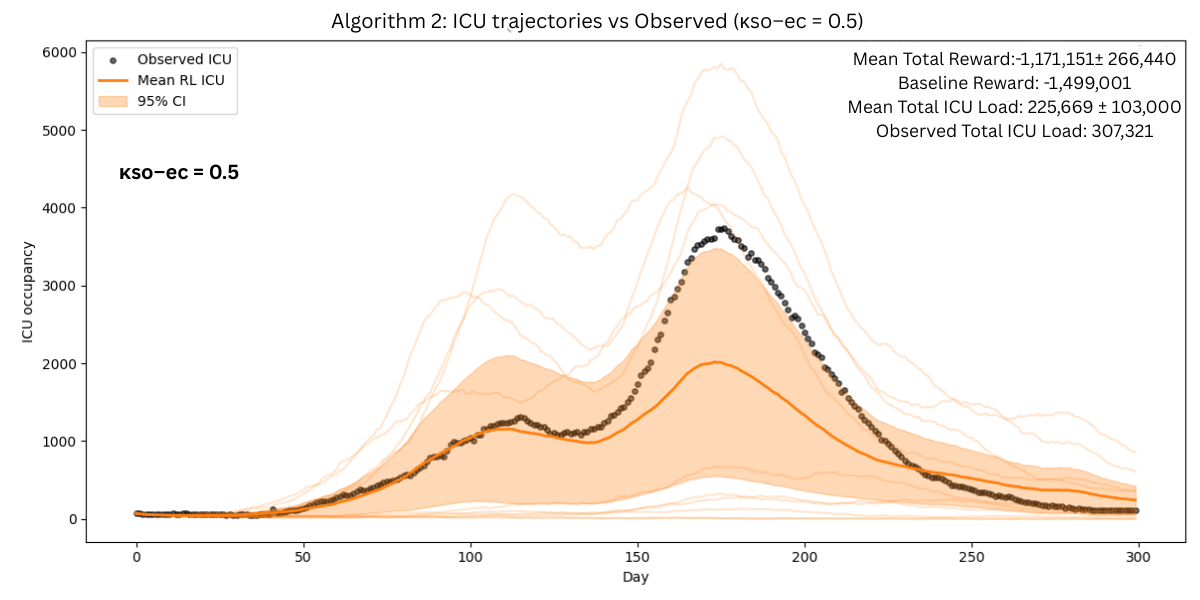}\\[1ex]
  \includegraphics[width=0.85\textwidth]{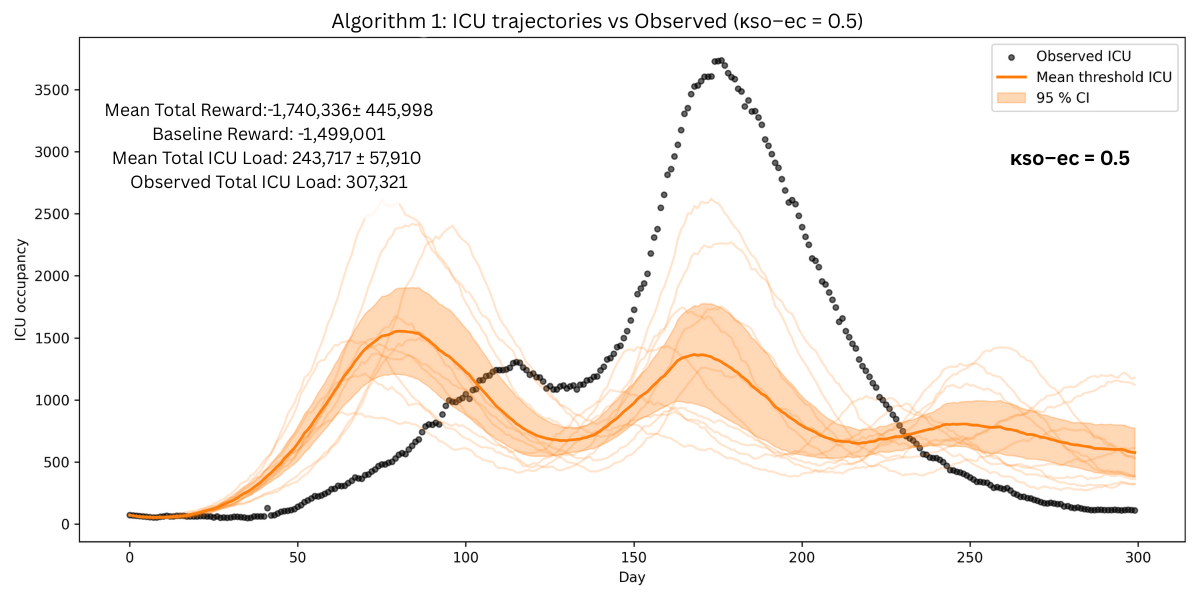}
  \caption{%
    Top Panel: ICU load trajectory under the $Q$‐learning controller with $\kappa_{\mathrm{so-ec}}=0.5$; 
    Bottom Paenl: ICU load trajectory under the threshold‐based controller with $\kappa_{\mathrm{so-ec}}=0.5$.
  }
  \label{fig:policy_05_comparison}
\end{figure*}

\begin{figure*}[!ht]
  \centering
  \includegraphics[width=0.85\textwidth]{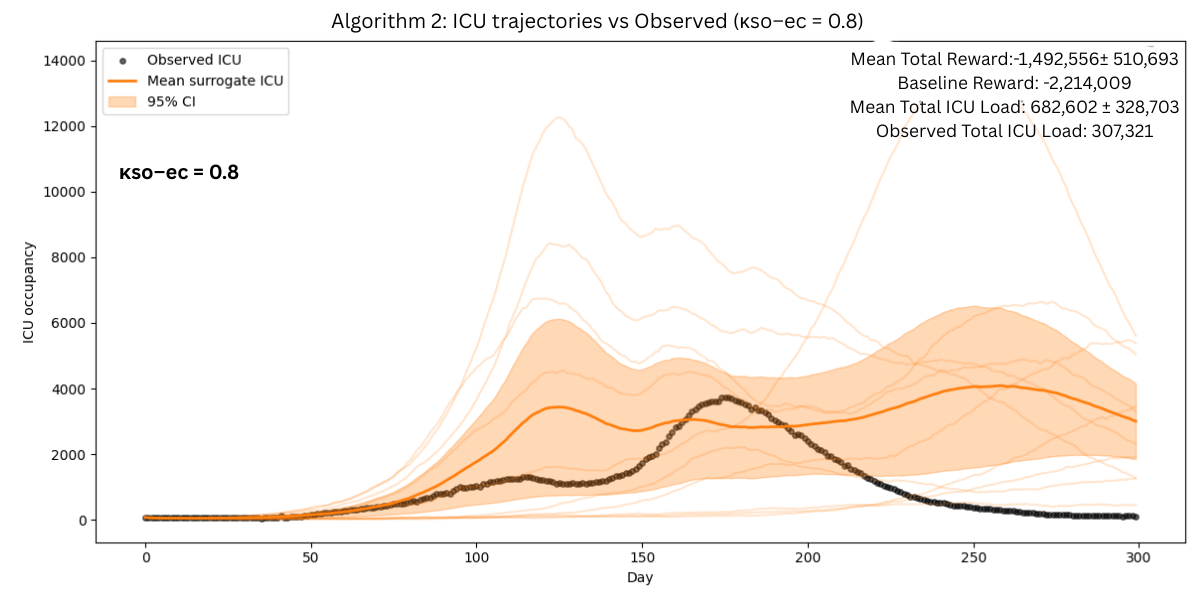}\\[1ex]
  \includegraphics[width=0.85\textwidth]{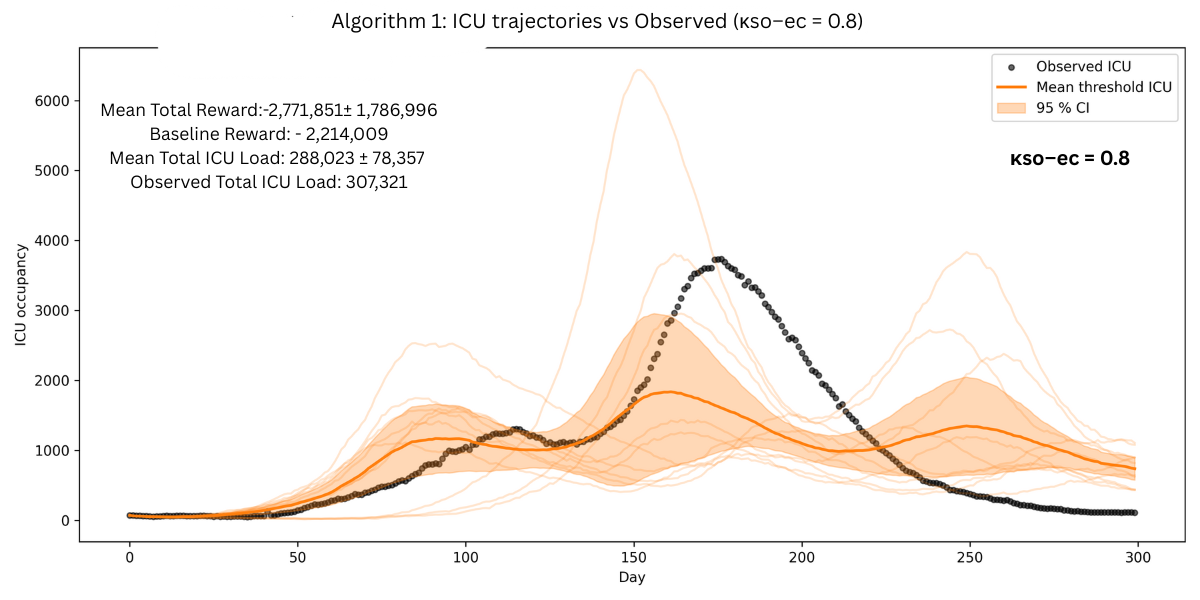}
  \caption{%
    Top Panel: ICU load trajectory under the $Q$‐learning controller with $\kappa_{\mathrm{so-ec}}=0.8$; 
    Bottom Panel: ICU load trajectory under the threshold‐based controller with $\kappa_{\mathrm{so-ec}}=0.8$.
  }
  \label{fig:policy_08_comparison}
\end{figure*}

We also ran both controllers with different cost scales, $\kappa_{\textrm{so--ec}} = 0.5$, $\kappa_{\textrm{so--ec}} = 0.8$, with the results shown in Figures \ref{fig:policy_05_comparison}--\ref{fig:policy_08_comparison}. 
As $\kappa_{\text{so--ec}}$ increases, the algorithms shift their focus towards a reduction in the socio-economic costs of  interventions. This leads to a tendency to avoid strict interventions, even if that results in higher total ICU burdens. 
For both higher cost settings, the $Q$-learning algorithm outperforms the threshold-based controller in terms of the total reward, while the corresponding rewards are very close when $\kappa_{\textrm{so--ec}} = 0.2$. For fairness both algorithms were run for the same amount of computational time. 
First, the superior overall performance of the $Q$-learning algorithm in our experiments could be attributed to the following factors.
First, the threshold-based policy leads to lower degree of flexibility against the $Q$-learner, even when the grouping of ICU values into categories used in the latter approach is taken under consideration. 
Then, the $Q$-learning approach is more efficient from a computational viewpoint, as it optimises the intervention strategy faster compared to the threshold controller due to the latter requiring more extensive Monte Carlo search roll-outs to evaluate different intervention strategies. 
In contrast, the $Q$-learning approach seems to achieve competitive or even superior performance in less time, making it a more computationally efficient solution, and one which  is particularly suited for large-scale or real-time applications.


Additionally, we observed that the $Q$-learning algorithm exhibits less uncertainty across the $10$ replications in terms of total reward compared to the threshold-based controller. This is likely due to the threshold-based controller's sensitivity to initial choices of thresholds in the first few blocks. Since the threshold-based controller does not re-run the full grid search at each decision block to save computational time, the initial threshold values affect the evolution of the ICU trajectory, leading to higher variability across replicates. In contrast, the $Q$-learning algorithm samples multiple parameter sets from the current posterior at each decision block, runs $Q$-learning to convergence for each sampled set, and then averages the resulting $Q$-values to select the action with the highest expected value. This procedure produces stable $Q$-values that translate into more consistent and reliable total rewards across replicates. This is a critical consideration for real-world applications that demand consistent decision-making.

In summary, as we increase the socio-economic cost, both algorithms adjust their strategies, with the threshold controller becoming less effective in maximising the total reward, especially at higher cost scales. The $Q$-learning algorithm, on the other hand, adapts more efficiently, leading to better performance, lower uncertainty, and higher total rewards, even as the cost of intervention rises.

\subsection{$Q$-Learning Convergence}
\label{sec:convergence_q}

\begin{figure*}[t]
  \centering
  \includegraphics[width=0.92\linewidth]{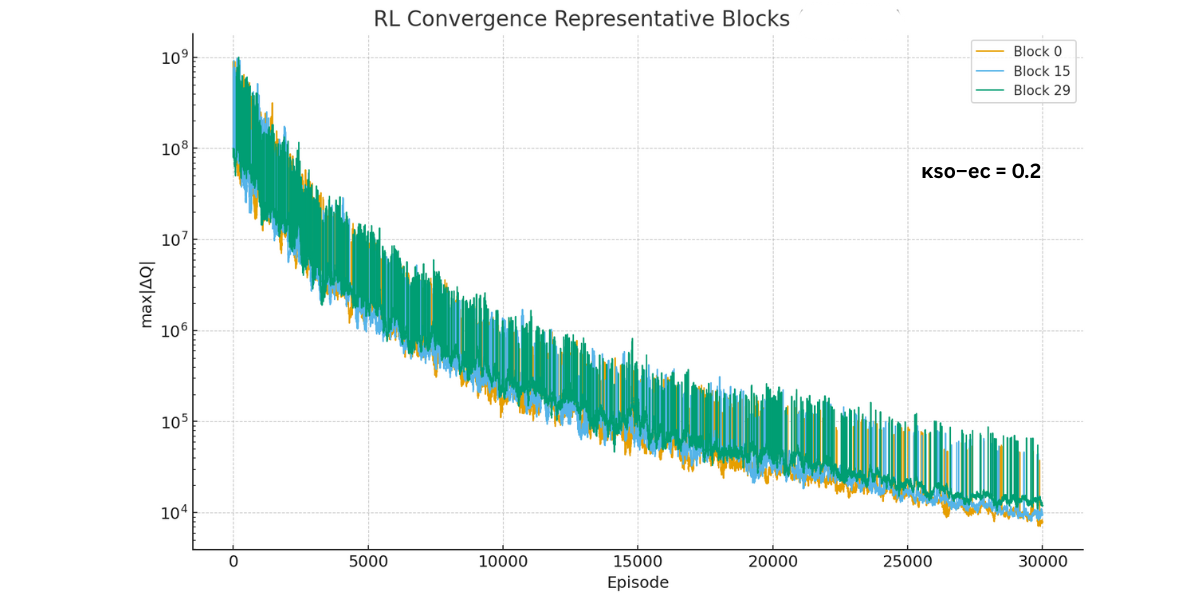}
  \caption{Convergence of $\bar Q$ within representative blocks under cost scale  
  $\kappa_{\textrm{so--ec}}=0.2$.
  The y-axis is $\max_{g,a}|\bar Q_e-\bar Q_{e-1}|$ (log-scale).
  Early episodes show larger, noisy updates; later episodes show a calm plateau.}
  \label{fig:q_convergence}
\end{figure*}

For the $Q$-learning-based Algorithm \ref{alg:smc2_post_avg_q_learning}, 
at each decision time $t_0=(b-1)\Delta$, 
we simulate $E$ planning episodes, each of length $H$.
At episode index $e$, we keep track of the posterior-averaged $Q$-table:
\begin{equation*}
\bar Q_e(g,a)=\tfrac{1}{K}\sum_{k=1}^K Q^{(k)}_e(g,a),
\end{equation*}
with each $Q^{(k)}_e$ trained under a $\theta^{(k)}\sim p(\theta|\mathcal{Y}_{t_0})$ provided from the SMC\textsuperscript{2} 
sampler that targets the posterior up to time $t_0$.
Figure \ref{fig:q_convergence} displays, on a log-scale, the episode-wise maximum absolute change evolution for three different decision times. I.e., Figure \ref{fig:q_convergence} keeps track of the following values, across all episodes $e$:
\begin{equation*}
\mathrm{max}|\Delta Q|(e)=\max_{g,a} \Big|\bar Q_e(g,a)-\bar Q_{e-1}(g,a)\Big|.
\end{equation*}
A rapid decay of $\mathrm{max}|\Delta Q|$ and a low-variance tail indicate that the Bayes-averaged value surface $\bar Q$ has stabilised.

The decay pattern follows directly from Algorithm~\ref{alg:smc2_post_avg_q_learning}:
(i) early episodes combine $\varepsilon$-greedy exploration with simulation noise from the SEIR--VU dynamics $s_{t+1}\sim p(s_{t+1}\,|\,s_t,a_t;\theta)$ and dispersion across the posterior draws $\{\theta^{(k)}\}_{k=1}^K$, producing relatively large and volatile TD updates;
(ii) as episodes accumulate, the steps become less variable and $\varepsilon$ decays, so updates shrink;
(iii) averaging over $K$ sample reduces the variance of the update direction, further damping oscillations.

\medskip
\textbf{Convergence Diagnostics.}
We adopt an operational stopping rule adapted from dynamic-programming practice. Within each decision block we monitor the sup-norm change of the posterior-averaged state-action value function,
$\mathrm{max}|\Delta Q|(e)=\max_{g,a}\lvert \bar Q_e(g,a)-\bar Q_{e-1}(g,a)\rvert$,
and declare convergence once this quantity falls below a small \emph{relative} tolerance (e.g., $10^{-5}$–$10^{-4}$ of the maximum absolute $Q$-differences over the first $2,000$ episodes), \emph{along with} a policy-stability check where the greedy policy induced by $\bar Q$ produces less than $1\%$ of different actions compared to the obtained $Q$-table at the previous episode.
This mirrors the classical $L_\infty$ stopping criterion used for value/policy iteration \cite{puterman2014markov,Bertsekas2012} and complements asymptotic convergence guarantees for tabular $Q$-learning \cite{WatkinsDayan1992,BertsekasTsitsiklis1996,SuttonBarto2018}. Thresholds and patience are application-level hyperparameters chosen for numerical robustness rather than theoretically mandated constants. (Equivalently, one may monitor an empirical Bellman residual, $\|\bar Q_{k+1} - \bar Q_{k}\|_\infty$, estimated via TD errors, an approach that follows the same spirit and is described in the same citations as above.)

From our results, we can conclude that SMC\textsuperscript{2}-guided posterior averaging makes $Q$-learning viable in our short, blockwise planning regime: despite early noise and block-to-block non-stationarity, $\bar Q$ reliably stabilises within each block before deployment, and the induced actions are stable.

\section{Conclusions}\label{conclusion}

This paper presents a real-time decision-support framework designed for epidemic control, leveraging a suitably tailored SEIR–VU model integrated with sequential Bayesian filtering techniques such as SMC\textsuperscript{2}. By incorporating RL, the framework dynamically adjusts intervention strategies in response to real-time ICU data, with the objective of minimising ICU burden while accounting for socio-economic costs. The approach provides a balanced solution that appropriately weights the immediate health impact of interventions versus their longer-term socio-economic consequences.

To address the challenge of quantifying socio-economic costs, a tailored method was developed through a combination of empirical simulations and consultations with epidimiological experts. This ensures that the cost function used in the framework reflects realistic socio-economic trade-offs, providing a contextual and data-driven approach to epidemic management. The framework’s adaptability makes it suitable for use in a variety of epidemic scenarios, offering flexibility beyond the COVID-19 pandemic.

Our approach was validated using real-world UK ICU time-series data. The real data were used to train the model generator, which was then employed to generate synthetic data under various intervention policies selected by the controllers. Two RL controllers were tested: one based on an interpretable ICU-threshold policy implemented through Monte Carlo grid search and another using a posterior-averaged $Q$-learning algorithm. These controllers were compared with actual government intervention strategies over a 300-day period during the COVID-19 pandemic.

Under scenarios where the designed socio-economic costs are calibrated to match the realised costs of the government NPIs (we calculated the governmental intervention cost by inserting the true deployed actions by the government in (\ref{eq:cost})), both RL controllers substantially reduced ICU burden relative to the government’s intervention. Notably, for higher socio-economic cost settings, the $Q$-learning controller consistently outperformed the threshold-based one in terms of both uncertainty and total reward, despite both controllers being provided with the same computational resources. The reason for this superior performance lies in the $Q$-learning algorithm’s ability to reduce uncertainty through posterior averaging and its efficient exploration of the state space. By leveraging past experiences and optimising future interventions dynamically, the $Q$-learning controller was able to provide more stable and reliable outcomes, while the threshold-based controller, although interpretable, was more sensitive to initial choices and variations in the threshold grid. 

The broad public health interpretation of the source of those gains suggest that one should intervene slightly earlier and also relax strict interventions somewhat earlier. Roughly, at comparable socio-economic cost, this timing shift reduces cumulative ICU burden and peak pressure by optimising the timing and scale of the interventions. This is not meant as a criticism to the decisions made at the time, but simply highlights the potential of combining real-time Bayesian learning with RL for optimised decision-making for epidemic control, especially when faced with the need to balance public health and economic concerns. Hence, our framework showed that real-time decision support could be an essential tool for policymakers, offering valuable insights for managing interventions efficiently.

We note that the realised actions by the authorities were based on input from multiple experts and were far from unsophisticated. 
The improved results reported in this work benefit from the fact that our policy learning is based upon the particular reward function we have chosen -- though we have tried to control for this factor by trying a number of values for $\kappa_{\textrm{so--ec}}$. Also, our framework does not restrict the choice of an action at a decision time, when realistic authority decisions might involve consideration of smooth transitions amongst actions of varying severity. Such a point can be studied in future research.

In conclusion, the proposed framework addresses epidemic control challenges of the type posed during the COVID-19 pandemic and provides a robust tool for future epidemic scenarios. By integrating real-time data, sequential learning, and explicit handling of parameter uncertainty, the framework can optimise 
intervention strategies across diverse contexts.

\medskip
\textbf{Future Work Directions.} 
Despite the promising results, several avenues for future work remain. One key aspect is the further refinement of the socio-economic cost function. While our current method is empirically derived and provides a practical tool for decision-making, it is not fully data-driven. Future work should focus on more accurate quantification of socio-economic costs, potentially through extensive collaboration with policymakers and experts in economics and public health. This would allow for a better understanding of the trade-offs between ICU burden and socio-economic impact, ensuring that the cost function aligns more closely with real-world dynamics.

Another direction for future research is exploring generalisations of our the framework to support (potentially discounted) long-term epidemic control beyond a fixed time horizon. This would require considering the manner in which intervention strategies evolve over extended periods, integrating more dynamic socio-economic cost models and refining the estimation of epidemic parameters as new data becomes available. Incorporating uncertainties regarding future interventions and their impacts would help enhance decision-making in highly uncertain environments.

Additionally, the proposed framework can be further enhanced by handling multiple pathogens such as influenza and respiratory syncytial virus, incorporate real-time data from surveillance pipelines, and accommodate a broader range of cost specifications. Extending the modularity of the framework would make it adaptable to a wide variety of epidemic control scenarios, ensuring that it remains a robust and flexible tool for policymakers facing emerging infectious diseases as well as regular surveillance periods.

Finally, it appears relatively straightforward, at least in principle, to extend the proposed method to multi-type epidemic models, for example by accounting for age-groups. This will likely require the incorporation of (possibly time-varying) contact matrices which are typically used in order to overcome potential identifiability problems. Extensions of this type can yield richer structures of control options, answering questions like: should interventions of variable severity be differentially applied to distinct age groups? This is not a trivial issue since those age groups are dynamically interacting over time and an answer to this question is the subject of current work.


\medskip
\paragraph{Code and data availability.}
The implementation of the SEIR--VU model, the threshold-based policy $\pi^\textrm{I}_\phi$ (Algorithm~\ref{alg:smc2_gridsearch}), and the posterior-averaged $Q$-learning controller $\pi^\textrm{II}$ (Algorithm~\ref{alg:smc2_post_avg_q_learning}), together with all scripts required to reproduce the experiments in Sections~6.2--6.3, are available from \url{https://github.com/giacomoiannucci01/seirvu-rl-epi-control}. The real ICU time series used in the paper is taken from NHS England's \emph{COVID-19 Hospital Activity} dataset (NHS England, ``COVID-19 Hospital Activity'', available from \url{https://www.england.nhs.uk/statistics/statistical-work-areas/covid-19-hospital-activity/}, accessed on 25th of June 2025).

\bibliography{sn-bibliography}

\end{document}